\documentclass[usenatbib]{mn2e}
\usepackage{times, enumerate}
\usepackage{epsfig}
\usepackage{amsmath}
\usepackage{amssymb}
\title[Varying IMF and the efficiency of galaxy formation]{The stellar mass function and efficiency of galaxy formation with a varying initial mass function}

\author[McGee et al.]{Sean L. McGee$^{1}$\thanks{Email:
    mcgee@strw.leidenuniv.nl}, Ryosuke Goto$^{1,2}$, Michael L. Balogh$^{1,3}$
 \\
$^{1}$Leiden Observatory, Leiden University, PO Box 9513, 2300 RA 
Leiden, The Netherlands\\ 
$^{2}$Department of Astronomy, Graduate School of Science, The University of Tokyo, 7-3-1 Hongo, Bunkyo-ku, Tokyo, 113-0033, Japan \\
$^{3}$Department of Physics and Astronomy, University of Waterloo, Waterloo, Ontario, N2L 3G1, Canada\\
}

\date{\today}
\def\LCDM{$\Lambda$CDM$~$}
\def\Mdoth{$h^{-1}~$M$_\odot$}

\def\Msun{M$_\odot$}

\def\Mpch{$h^{-1}~$Mpc}

\def\sigmaeight{$\sigma_8$}
\def\OmegaM{$\Omega_\mathrm{M}$}
\def\kms{${\rm km\ }{\rm s}^{-1}$}
\def\kmsmpc{\>{\rm km}\,{\rm s}^{-1}\,{\rm Mpc}^{-1}}

\def\Mstellar{$M_\odot~$}

\def\Halpha{H$\alpha$}

\def\Chandra{{\it Chandra}}

\def\petromag{\textsc{petroMag}}
\def\modelmag{\textsc{modelMag}}

\def\beq{\begin{equation}}
\def\eeq{\end{equation}}
\def\lowz{0.005}
\def\highz{0.20}

\begin{document}

\maketitle
\begin{abstract}

Several recent observational studies have concluded that the initial mass function (IMF) of stars varies systematically with galaxy properties such as velocity dispersion. In this paper, we investigate the effect of linking the circular velocity of galaxies, as determined from the Fundamental Plane and Tully-Fisher relations, to the slope of the IMF with parameterizations guided by several of these studies. For each empirical relation, we generate stellar masses of $\sim$ 600,000 SDSS galaxies at z$\sim$0.1, by fitting the optical photometry to large suites of synthetic stellar populations that sample the full range of galaxy parameters. We generate stellar mass functions and examine the stellar-to-halo mass relations using sub-halo abundance matching. At the massive end, the stellar mass functions become a power law, instead of the familiar exponential decline. As a result, it is a generic feature of these models that the central galaxy stellar-to-halo mass relation is significantly flatter at high masses (slope $\sim -0.3$ to $-0.4$) than in the case of a universal IMF (slope $\sim -0.6$). We find that regardless of whether the IMF varies systematically in all galaxies or just early types, there is still a well-defined peak in the central stellar-to-halo mass ratio at halo masses of $\sim$ 10$^{12}$ \Msun.  In general, the IMF variations explored here lead to significantly higher integrated stellar densities if the assumed dependence on circular velocity applies to all galaxies, including late-types; in fact the more extreme cases can be ruled out, as they imply an unphysical situation in which the stellar fraction exceeds the universal baryon fraction. 

\end{abstract}

\begin{keywords}
galaxies: evolution, galaxies: formation
\end{keywords}

\section{Introduction}

The collapse and fragmentation of gas into stars results in a stellar mass distribution known as the initial mass function (IMF). Unfortunately, there is no first principles consensus about the origin of this distribution or how it might depend on local conditions. Consequently, progress is driven by empirical results, such as the direct determination of the distribution via star counts as pioneered by \citet{salpeter_imf}, who found the massive end of the IMF was well modeled by a declining power law. Subsequent studies have shown that, below a solar mass, there are fewer stars than an extrapolated power law would suggest. The exact form of this turnover, and the resulting distribution within the Milky Way is a matter of some debate, resulting in several different parameterizations \citep[eg.,][]{salpeter_imf, Miller_scalo, kennicutt+83, scalo, kroupa, chabrier_imf}. Despite this range of results, it appears that the majority of nearby measurements of the IMF are consistent with a single, universal form \citep{Bastian_review}. However, there is an increasing range of observational indications from distant galaxies, and solid theoretical motivation, to suggest the IMF may be altered depending on the local conditions under which star formation occurs.

There are several possible explanations for how and why gas fragments as it does and most predict variations in the IMF depending on local conditions --- be it redshift, metallicity, etc. The explanation with the longest history is that the collapse occurs when the gas self-gravity becomes stronger than the thermal pressure, eg. the Jeans instability \citep{Jeans_instability}. As this is directly related to the temperature of the gas, it suggests variations in the Jeans mass will potentially depend on, at least, metallicity (due to its effect on cooling rates), redshift (due to the higher cosmic background temperature) and environment (due to the variation in the external heating rate) \citep{Larson1998, Larson2005}. In an alternative explanation, the mass of stars is set by a competition between accretion rate, which depends on the sound speed, and stellar outflows, which are more effective in high metallicity gas due to the ease of coupling to radiation \citep{silk_imf, adams_fatuzzo_1996}. Similar variations are expected if the IMF is set by the turbulence in the ISM \citep{padoan_nordlund, Hennebelle_chabrier_1998, Hopkins_ISM_IMF}. There is also a semi-empirical framework, the integrated-galactic IMF (IGIMF), which suggests the IMF varies with local cloud density and metallicity \citep[eg.,][]{Kroupa_review_2013}. On the other hand, there are also arguments that radiative feedback self-regulates star formation resulting in a nearly universal IMF \citep{Bate2009, Myers_imf_2011}.

Among the first observational indications for a varying IMF came from a study by \citet{Vazdekis_1996} which showed that the ($V$ - $K$) - Mg$_2$ relation of massive elliptical galaxies could not be reproduced with a constant IMF. Subsequently, \citet{Cenarro_2003} found that an observered anti-correlation between calcium triplet abundance and velocity dispersion in early type galaxies could be indicative of a varying IMF, while similar results were seen in bulges of late type galaxies \citep{Falcon-Barroso_2003}. These early indications were very intriguing, but awaited further confirmation from other IMF probes.

Excitingly, in the past few years, there are now several lines of observational evidence which strength the case for IMF variations in galaxies outside the Milky Way. Among the first of these recent indications came from comparing a galaxy's total mass --- either through dynamics, strong gravitational lensing, or a combination of the two --- to the stellar mass implied by modeling its stellar populations. \citet{Treu_2010} showed that the total central mass of a sample of early type galaxies with strong gravitational lenses was significantly higher than that implied by the observed stellar light if the IMF has the form advocated by \citet{chabrier_imf}. This, so called `IMF mismatch', was found to nearly disappear if a Salpeter IMF was assumed. Intriguingly, the size of this IMF mismatch appeared to vary systematically with the velocity dispersion of the system. However, this mismatch might also be caused by a variation in the density profile of dark matter at the center of the galaxies. The results require either a non-universal IMF or a non-universal dark matter profile, or a combination of the two. The need for some additional source of mass was strengthened with larger samples and with analysis which allowed for physical models of the adiabatic contraction of dark matter halos \citep{Auger_2010_expanded, Auger_2010_contraction}.

The incompatibility of a universal IMF and universal halo contraction is also required to reconcile galaxy circular velocities derived from the Faber-Jackson and Tully-Fisher relations and the stellar mass of the galaxy \citep{Dutton_2010_imf}. Studies of very compact galaxies, in which the stellar mass dominates the total internal mass, allow this degeneracy to be broken, and indicate strong evidence for IMF variation \citep{Dutton_2012_compact}. Similarly, detailed dynamical constraints enabled by integral field spectroscopy of large samples of early type galaxies also find that the stellar mass to light ratio must systematically vary with velocity dispersion \citep{Cappellari_2012, Cappellari_2013}. Unfortunately, these probes are only sensitive to the total mass in stars and not the exact form of the IMF. Crucially, the old stellar populations typical of early type galaxies can have higher mass to light ratios (M/L) than a Chabrier stellar population with either a dwarf-enhanced (bottom-heavy) or dwarf-deficient (top-heavy) IMF. This seeming paradox arises because old populations of top-heavy distributions have a large fraction of their mass in non-radiating stellar remnants. 

Concurrently with the lensing and dynamical evidence for IMF variation, improvements in empirical libraries of near infrared stellar spectra \citep{Rayner_2009} and red sensitive CCDs have led to probing the IMF through direct measurement of absorption lines which are prevalent only in dwarf stars. \citet{vanDokkum2010} showed that the Na I doublet and Wing-Ford molecular FeH band, both features strong in dwarfs and non-existent in massive stars, are well detected in massive, cluster ellipticals. The detection of dwarf sensitive line indices gives evidence that galaxies become more `bottom-heavy' with increased velocity dispersion, rather than the otherwise dynamically allowed `top-heavy'. While the direct measurements of such line indices are very subtle (eg. 1-3 $\%$ variations in the spectrum), a comparison with observed spectra of globular clusters of similar age, metallicity and alpha enhancement shows such clusters do not have dwarf-enhanced populations, as they must given their very low mass to light ratios \citep{vanDokkum_globular_2012}. Larger samples of early type galaxies have shown that these direct line measurements also imply the IMF varies with velocity dispersion \citep{VDC_paper1_2012, VDC_paper2_2012}. Along with the increase of dwarf-sensitive indices with velocity dispersion, encouragingly, the data also show a decrease in the strength of giant-sensitive Ca II indices \citep{VDC_paper1_2012, VDC_paper2_2012} and are unlikely to be due to further elemental variation in galaxies \citep{Conroy_vD_elements, VDC_paper2_2012}.

There has been considerable effort in finding IMF-sensitive line indices in the more easily accessible optical region where huge samples of spectra are available. Several of these lines have been identified and have been applied to extensive samples of stacked spectra to obtain tight constraints on IMF variation with velocity dispersion. In particular, \citet{ferreras_2013_imf}, \citet{LaBarbera_2013} and \citet{spiniello2013} have analyzed stacked spectra of SDSS early type galaxies and obtain relations between IMF slope and velocity dispersion. While these results have relatively small statistical errors, they are still potentially dominated by systematics. As an important test, in some cases lensing/dynamical determinations of the IMF mismatch and detailed spectral indices analysis have been done on the same systems independently and generally been found to give consistent results \citep{tortora_2013, Barnabe_2013, Conroy_compact_2013} although there are well measured exceptions \citep{Smith_lens_2013}. 

Despite the existence of such strong trends with velocity dispersion, it is important to remember that many galaxy observables are significantly correlated. Indeed it has been suggested that the IMF varies more strongly with Mg/Fe than velocity dispersion, and likely indicates the IMF varies with star formation mode rather than system mass \citep{SmithLucey2012}. It has also been suggested, as part of the integrated-galaxy IMF (IGIMF) theory, that the $M/L$ results are a natural outcome of a IMF which depends on density and metallicity of the birth clouds \citep{weidner_2013_IGIMF}. Nonetheless, the existence of such relations with velocity dispersion allows the implications to be easily extended to all systems.

Until now, our discussion of observational indications of a varying IMF have been restricted to relatively massive, early type galaxies. However, there are several reasons to expect that the IMF in late type and/or low mass galaxies varies in a similar way. First, the principal reason previous studies were restricted to early type galaxies is because their old stellar populations remove the degeneracies generated by the uncertain star formation history of the galaxies, as well as allowing absorption lines to be measured without contamination. However, there are still methods that could measure the IMF in newly formed stellar populations. \citet{Hoversten_glazebrook} used the ratio of \Halpha\ emission to $g$-$r$ color as a lever arm to measure the massive end slope of the IMF in SDSS star forming galaxies. They found that, while bright star forming galaxies have a Salpeter-like slope, fainter galaxies prefer steeper slopes. Unfortunately, this method presents a degeneracy between the maximum mass of stars formed and the IMF slope, suggesting that a natural explanation for the steeper slopes in fainter galaxies is that such galaxies have lower mass giant molecular clouds (GMCs) and thus form fewer very massive stars. \citet{Gunawardhana_imf} used the same method in GAMA galaxies and found similar results, but suggested that the primary driver of IMF variation was the star formation rate of the galaxy.

It has been suggested that a varying IMF is needed to explain the high number and luminosity of sub-mm detected galaxies \citep{Baugh_05}, and this tension remains even though some of the most luminous sub-mm sources have contributions from multiple galaxies \citep{karim_alma, hodge_alma}. A similar varying IMF was suggested to reconcile the apparent discrepancy of the stellar density of the Universe with its inferred integrated star formation history assuming a universal IMF \citep{Wilkins_2008_imf, Wilkins_2008, Kang_imf}, although recent homogeneous analysis has suggested there is no discrepancy even with a universal Chabrier IMF \citep{Sobral_2013, Behroozi_2013}. 

Despite these diverse, recent results that show the IMF varies systematically within the galaxy population, most statistical studies of galaxies still assume a universal IMF. It is the goal of this paper to use the best empirical results to impose an IMF on a galaxy depending on its circular velocity. We then calculate stellar masses through spectral energy distribution fitting to a large grid of models generated with the given IMF. The stellar mass functions are then compiled and compared to the dark matter subhalo mass function to find the efficiency of galaxy formation under the assumption of subhalo abundance matching. We stress that our goal in this paper is not to determine the IMF, but rather explore the implications of previous studies.

The paper is organized as follows. We discuss the data used in the paper in \textsection \ref{data}, the form of our IMF variations in \textsection \ref{mod-imf}, our stellar population synthesis and SED fitting in \textsection \ref{stelpop}. We present the stellar mass and subhalo matching results in \textsection \ref{results}, discuss the results in \textsection \ref{discuss} and present our conclusions in \textsection \ref{conclusions} . We adopt a \LCDM cosmology with the parameters as determined by the Wilkinson Microwave Anisotropy Probe (WMAP) after seven years of data (WMAP7); namely $\Omega_{\rm m} = 0.27$, $\Omega_{\Lambda}=0.73$,  $h=H_0/(100 \kmsmpc)=0.70$ and $\sigma_8$ = 0.81 \citep{komatsu_wmap7}. All the stellar masses derived from our SED fitting are based on observed fluxes, and thus converting to different $H_0$ introduces two factors, whereas the subhalo mass function is based on dynamics and thus only one factor of $H_0$ \citep[eg.,][]{Croton_littleh}. All magnitudes are stated within the AB magnitude system \citep{Oke+83}.

\section{Data}\label{data}

Our goal is to derive stellar mass functions in the local Universe using empirical relations to connect the IMF to the circular velocity of the galaxy. As such, we require a large spectroscopic redshift survey with a well understood selection and a large complement of derived galaxy properties. The best survey for this purpose is the main galaxy sample of the Sloan Digital Sky Survey \citep[SDSS;][]{yorkSDSS, SDSSmainsample}.

\subsection{SDSS photometry and derived measures}

Our base catalog is the seventh data release of the SDSS \citep{sdss_dr7} as presented in the New York University Value-Added Galaxy Catalog \citep[NYU-VAGC;][]{NYU_VAGC}. The main galaxy survey targets galaxies with $r_{\mathrm{petro}}$ $<$ 17.77 after a correction for galactic extinction derived from the dust maps of \citet{dustmaps}. Additional cuts based on the observed light profile were made to efficiently separate galaxies from stars and to avoid very low surface brightness galaxies. While each of these cuts has the potential to induce selection effects in the reconstructed stellar mass function, in the mass range we are interested in, good agreement is found with surveys that have less aggressive (and less efficient) selection criteria \citep[eg.][]{baldry_gama_stelmass}. We restrict our sample to galaxies targeted as part of this main galaxy survey and with redshifts between z = 0.005 and z=0.20, which results in a sample of 635,108 galaxies spread across 7966 square degrees.

The SDSS photometric pipeline measures magnitudes and fluxes in each of the $ugriz$ bands using several different methods -- two of which are used in this paper (\petromag\ and \modelmag). \petromag\ magnitudes were used in the the original galaxy selection, and are measured within a circular aperture to a given fraction of the Petrosian radius, as described in \citep{blanton_earlylumfun}. The \modelmag\ magnitudes are measured by fitting radial profiles to the galaxy (either deVaucouleurs or exponential) and integrating the function to a given radius. \modelmag\ and \petromag\ should be consistent for most galaxies, although deviations are seen in the fluxes of massive galaxies and thus their calculated stellar masses \citep{Bernardi_2013, He_photometry_massive}.

\subsubsection{Inferred velocity dispersions and circular velocities}

The SDSS spectroscopic pipeline measures velocity dispersions for each target classified as a galaxy by directly comparing the spectrum to template spectra convolved with a range of velocity dispersions. The instrumental resolution of the observed spectra are $\sim$ 70 \kms\ and the maximum velocity dispersion used to convolve the templates is 420 \kms, which sets lower and upper limits on the reliability of the measured velocity dispersions. Velocity dispersions measured in this way are straightforward to interpret for early type galaxies; however, for late type galaxies the velocity dispersion varies significantly with viewing angle and radius. For this reason, but also because of the finite range of well measured velocity dispersions, we will use velocity dispersions inferred from photometric properties of the galaxies as calibrated by the Tully-Fisher \citep[TF;][]{Tully_fisher} and Fundamental Plane \citep[FP;][]{Dressler87, Djorgovski_davis} relations. 

Velocity dispersions have often been inferred using a relation between the galaxy's size and its mass \citep[eg.,][]{bezanson_11}. However, for our purposes, this would require an iterative process as the velocity dispersion is linked to the IMF and thus the stellar mass. Given the vagaries in degeneracies of stellar population modelling, a robust convergence is unlikely. Thus, we will determine the velocity dispersions in a manner similar to \citet{desai_2004} and \citet{abramson_2013} using calibrations based on TF and FP relations. Of course, each of these calibrations is only usefully applied to late and early type galaxies respectively. Our classification of a galaxy as a late or early type galaxy is based on the best-fit Sersic index of the one component fit to the galaxy light profile. These measurements are provided by the NYU value added galaxy catalog and detailed in the appendix of \citet{blanton05}. Despite the relative simplicity and crudeness of this measurement, it is robust and generally agrees with more complicated measurements \citep{Graham_2001, blanton_review, guo_size_2009, Mosleh_2013}. We adopt the standard criteria that galaxies with $n$ $>$ 2.5 are early type \citep[eg.,][]{shen_size, patel_uvj_2011}.

The FP relation can be inverted to give an early type velocity dispersion, $\sigma_{\mathrm{early}}$, as such
\beq
\mathrm{Log}_{10} \sigma_{\mathrm{early}} = c_1 \  \mathrm{Log}_{10} R_e + c_2\  \mathrm{Log}_{10} I_0 + c_3
\eeq
where $R_e$ is the inclination-corrected half-light radius in kpc $\big($= $R_{50}$ $\sqrt{b/a}\big)$. The empirical coefficients, $c_i$, are determined by direct comparison to the measured velocity dispersion in the well-measured region (100 \kms $<$ $\sigma$ $<$ 400 \kms). From this, we adopt the values $c_1$ = 0.57 , $c_2$ = 0.39, and $c_3$ = 5.05. $I_o$ is the average surface brightness within the half-light radius and is calculated as
\beq
 \mathrm{Log}_{10} I_0 = -0.4  [m_r + 2.5 \mathrm{Log}_{10} (2 \pi R^2_0) - 10 \mathrm{Log}_{10}(1 + z)]
\eeq
where $m_r$ is the apparent r-band magnitude, $R_0$ is the apparent half-light radius in arcsec and z is redshift. Comparing the velocity dispersions determined in this way to the well measured spectroscopic measures for early type galaxies leads to a negligible offset ($\Delta$ = 0.42 \kms) and a small dispersion ($\sigma$ = 28.5 \kms). We convert this calculated velocity dispersion to a circular velocity assuming $v_{c}$ = $\sqrt{2}\ \sigma_{\mathrm{early}}$, which is expected for a singular isothermal sphere with a Maxwellian distribution of velocities \citep{Binney_tremaine_ed2}. The conversions between $v_C$ and $\sigma_{\mathrm{early}}$ used in the literature range approximately between $\sqrt{2}$ and $\sqrt{3}$ \citep[eg.,][]{desai_2004, cappellari_dynamics_2013, abramson_2013}. For our purposes, using a larger value would lead to higher IMF slopes in early type galaxies, and thus more extreme stellar masses.

In a similar manner, we can use the TF relation to determine the characteristic velocity for late type galaxies. \citet{pizagno_2007} used long-slit observations of a subset of SDSS galaxies to determine the TF relation for various determinations of the velocity. Here we will use the determination linking the $r$-band magnitude to the velocity at 2.2 scale lengths ($v_{2.2}$), which is equal to the circular velocity, $v_{\mathrm{c}}$ as such
\beq
\mathrm{Log}_{10} v_{\mathrm{c}} = \mathrm{Log}_{10} v_{2.2} = 2.192 - 0.140 (M_r^{\mathrm{corr}} + 21.107)
\eeq
where $M_r^{\mathrm{corr}}$ is the absolute $r$ magnitude k-corrected to z=0. We do not apply an internal dust correction. Combining the circular velocity determined from the TF relation for late type galaxies with that from the FP for early type galaxies gives a velocity characteristic of each galaxy regardless of type. This can be used to determine a relationship to the IMF.

\subsubsection{Vmax and constructing stellar mass functions} \label{sec-vmax}

As discussed above, the SDSS main galaxy sample is flux and surface-brightness limited. Many galaxies in the sample are only observable for a fraction of the full volume, so a correction must be applied. Assuming that evolution within the redshift range of the sample is negligible, then the corrections are straightforward and have been discussed by many authors for the SDSS sample. In detail, we follow the example of \citet{shen_size} and \citet{simard_sdss}, who calculate the observable volume of a galaxy given the minimum and maximum $r$ magnitudes ($r_\mathrm{min}$, $r_\mathrm{max}$) and the surface brightness limit. Given a galaxy with a measured redshift, $z$, and corresponding luminosity distance, $d_L(z)$, the maximum and minimum observable luminosity distances ($d_\mathrm{L,max}$, $d_\mathrm{L,min}$) are given as 
\beq
d_L(z_{max:r}) = d_L(z)10^{-0.2(r - r_\mathrm{max})} 
\eeq
and 
\beq
d_L(z_{min:r}) = d_L(z)10^{-0.2(r - r_\mathrm{min})} 
\eeq
where the $r$ band magnitudes have been corrected for galactic extinction.

We take the limiting surface brightness of the survey to be $\mu_{r,\mathrm{lim}}$ = 23 mag arcsec$^{-2}$, as beyond this limit the selection is more complicated. Given an observed surface brightness, $\mu_r$, the maximum observable redshift of the the galaxy is
\beq
z_{max:\mu_r} = (1 + z) 10^{(23 - \mu_{r,\mathrm{lim}})/10} - 1
\eeq

We impose hard limits on the maximimum (z$_{max:z}$ = \highz) and minimum (z$_{min:z}$ = \lowz) redshift of the sample to avoid evolution effects and the large angular size of very nearby galaxies. Thus each galaxies has a maximum and minimum redshift of observability: 
\beq
z_{min} = \mathrm{max}(\mathrm{z}_{min:r}, \mathrm{z}_{min:z})
\eeq 
\beq
z_{max} = \mathrm{min}(\mathrm{z}_{max:r}, \mathrm{z}_{max:\mu}, \mathrm{z}_{max:z})
\eeq
which set the limits to determine the maximum observable volume
\beq
V_{max} = \frac{1}{4\pi} \int d\Omega f(\theta, \psi) \int^{z_{max}}_{z_{min}} \frac{d^2_A(z)}{H(z)(1+z)} c dz
\eeq
where $\theta$ and $\phi$ give the position on the sky and $\Omega$ is the solid angle. $f(\theta, \psi)$ is the relative sampling fraction, $d_A$ is the angular distance, $H(z)$ is the Hubble parameter and $c$ is the speed of light. Given this $V_{max}$, we can calculate the stellar mass function by summing the inverse volume for each galaxy as such
\beq
\Phi_{\mathrm{Log_{10}M}} = \frac{1}{\Delta \mathrm{Log_{10} M}} \sum\limits_i \frac{1}{V_{max,i}} w_i
\eeq
where $\Delta \mathrm{Log_{10} M}$ is the width of the logarithmic mass bin and $w_i$ is any other galaxy weighting. $w_i$ can be used to add a correction for the large scale structure fluctuations which particularly effect very low redshift galaxies, and thus preferentially low mass galaxies \citep{Baldry_08}. In contrast to \citet{Baldry_08}, we only calculate the stellar mass function to 10$^{9}$ \Mstellar. As such, although we include this correction by calculating the number density in sliding magnitude bins following Baldry et al., our results are insensitive to this correction at the masses of interest.

\subsection{Dark matter subhalo function}

In order to gain physical intuition about the effect of a varying IMF on the galaxy population, we will relate the stellar mass function to the dark matter mass function using a direct one-to-one correspondence. This approach will allow us to examine the `efficiency' of galaxy formation -- or the ratio of stellar mass to halo mass --  in the later sections of the paper. For this task, we require a precise measurement of the halo (subhalo) mass function over a wide range of masses.  We will use the large scale, dark matter Millennium \citep[MS][]{MillSim} and the Millennium II \citep[MS-II][]{MillSimII} simulations. These simulations follow only the effects of gravity on the matter distribution with collisionless particles and thus represent an approximation to the growth of structure and halos in the universe. The original MS is a box with a one-sided length of 500 \Mpch\ with 2160$^3$ particles, while the MS-II has the same number of particles in a box of 100 \Mpch\ per side. As a result MS-II has a better mass resolution, which when combined with the MS allows a well resolved subhalo distribution across several orders of magnitude. 

These simulations require a specification of the cosmological parameters, which for the Millennium simulations were based on the early, year 1 results from the WMAP satellite \citep{spergel_wmap1} and thus have slightly different cosmological parameters to those determined with further analysis \citep{komatsu_wmap7} or more recent experiments \citep{planck_cosmo}. However, \citet{angulo_white} have show that by rescaling the simulations appropriately they can mimic large scale simulations in cosmologies with other parameters. We use the subhalo catalogs they have produced which rescale the Millennium I and Millennium II simulations to a WMAP7 cosmology \citep{komatsu_wmap7}, which we call MS-WMAP7 and MSII-WMAP7. The principal difference between these cosmologies is in \sigmaeight, which is 0.9 in WMAP1 and 0.807 in WMAP7, and \OmegaM\ which is 0.25 in WMAP1 and 0.272 in WMAP7.

%%%%%%%%%%%%%%%%%%%%%%%%%%%%%%%%%%%%%%%%%%%%%%%%%%%%%%%%%%%%%%%
\begin{figure}
\leavevmode \epsfysize=8.6cm \epsfbox{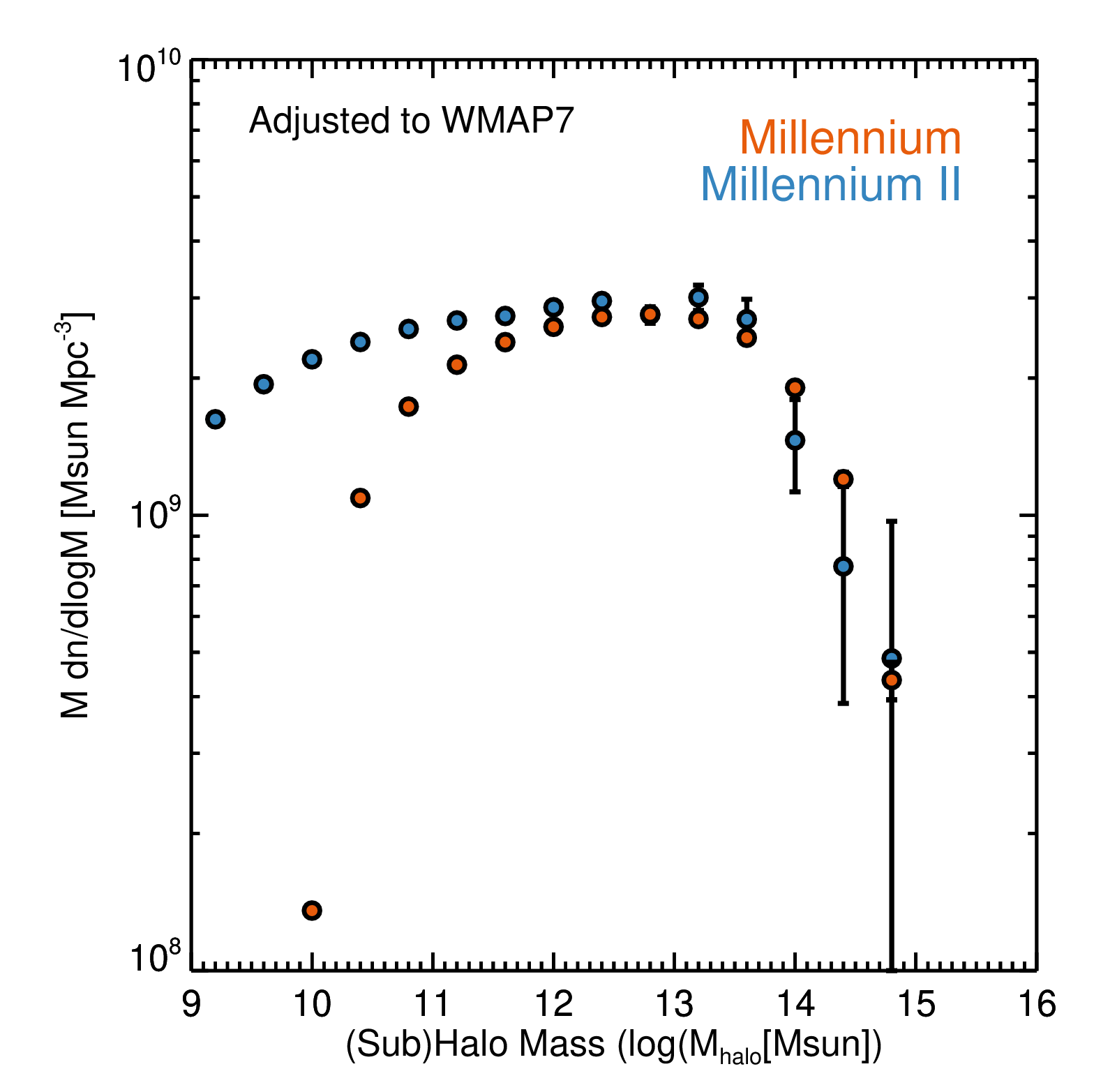}
\caption{The halo mass function at z=0 from the Millennium \citep{MillSim} and Millennium-II simulations \citep{MillSimII} after rescaling to a WMAP7 cosmology \citep{angulo_white}. This includes both main halos and satellite subhaloes within those main halos. The mass is taken to be the peak mass the subhalo has ever obtained. The MS-I results are used above 10$^{13}$ \Msun where the larger volume means the counts are not dominated by Poisson error. Below 10$^{13}$ \Msun the better mass resolution of MS-II makes it more robust.}
\label{fig_subhalo}
\end{figure}
%%%%%%%%%%%%%%%%%%%%%%%%%%%%%%%%%%%%%%%%%%%%%%%%%%%%%%%%%%%%%%%%

In Figure \ref{fig_subhalo} we show the halo and subhalo mass distribution in each of the MS-WMAP7(orange circles) and MSII-WMAP7 (blue circles) simulations. This plot is similar to Figure 1 in \citet{guo_2011_efficiency}, which used the original MS and MSII catalogs. Above a halo mass of 10$^{13}$ \Mdoth\ the smaller volume of MSII-WMAP7 leads to large uncertainties in the cosmological distribution of halos. Similarly, because of the relatively poor mass resolution of the MS-WMAP7, the distribution determined from it is systematically lower than the MSII-WMAP7 below 10$^{13}$ \Mdoth\ due to unresolved halos. For these reasons, we use the subhalo function determined from MS-WMAP7 above 10$^{13}$ \Mdoth\ and MSII-WMAP7 below this limit. As can be seen from the figure, MS-WMAP7 and MSII-WMAP7 agree well at 10$^{13}$ \Mdoth\ and thus there is no systematic discontinuity at that point.

\section{Empirical model of IMF variation}\label{mod-imf}

As discussed in the Introduction, there is no consensus about how star formation occurs in the Universe, but most models predict that the IMF should vary depending on local conditions. It is likely that this would result in IMF variations with many galaxy observables including redshift, metallicity, stellar or gas density, velocity dispersion, star formation `mode' (eg. bursts) and environment. Any observable in galaxy evolution is likely to have significant co-variances with other observables such that a relation in metallicity may actually reflect a more fundamental relation in redshift, star formation mode, etc. As such, without a realistic model of how the IMF changes with each of these variables, we adopt an empirical approach which relates the IMF variation in a galaxy to only the circular velocity, and thus the underlying potential well. Any physical model of IMF variation will ultimately have to reproduce these empirical results. For this investigation, we look at the implications of assuming these relations rather than attempting to reproduce them.

The majority of the evidence for systematic IMF variation come from observations of early type galaxies. However, there is no clear reason why these variations should only occur in early types. Indeed, in a hierarchical universe a significant fraction of a galaxy's stellar mass is accreted through consuming smaller galaxies of all types. Further, a large fraction of the passive, early type galaxies have been transformed from star forming, late type galaxies even since z $\sim$ 1 \citep{faber, bell, Muzzin13}. Therefore, it is not unreasonable to assume that a consistent variation of the IMF is seen in both early and late type galaxies. Nonetheless, we will present stellar mass functions where the IMF varies only for early type galaxies as well as all galaxies. 

%%%%%%%%%%%%%%%%%%%%%%%%%%%%%%%%%%%%%%%%%%%%%%%%%%%%%%%%%%%%%%%
\begin{figure}
\leavevmode \epsfysize=8.6cm \epsfbox{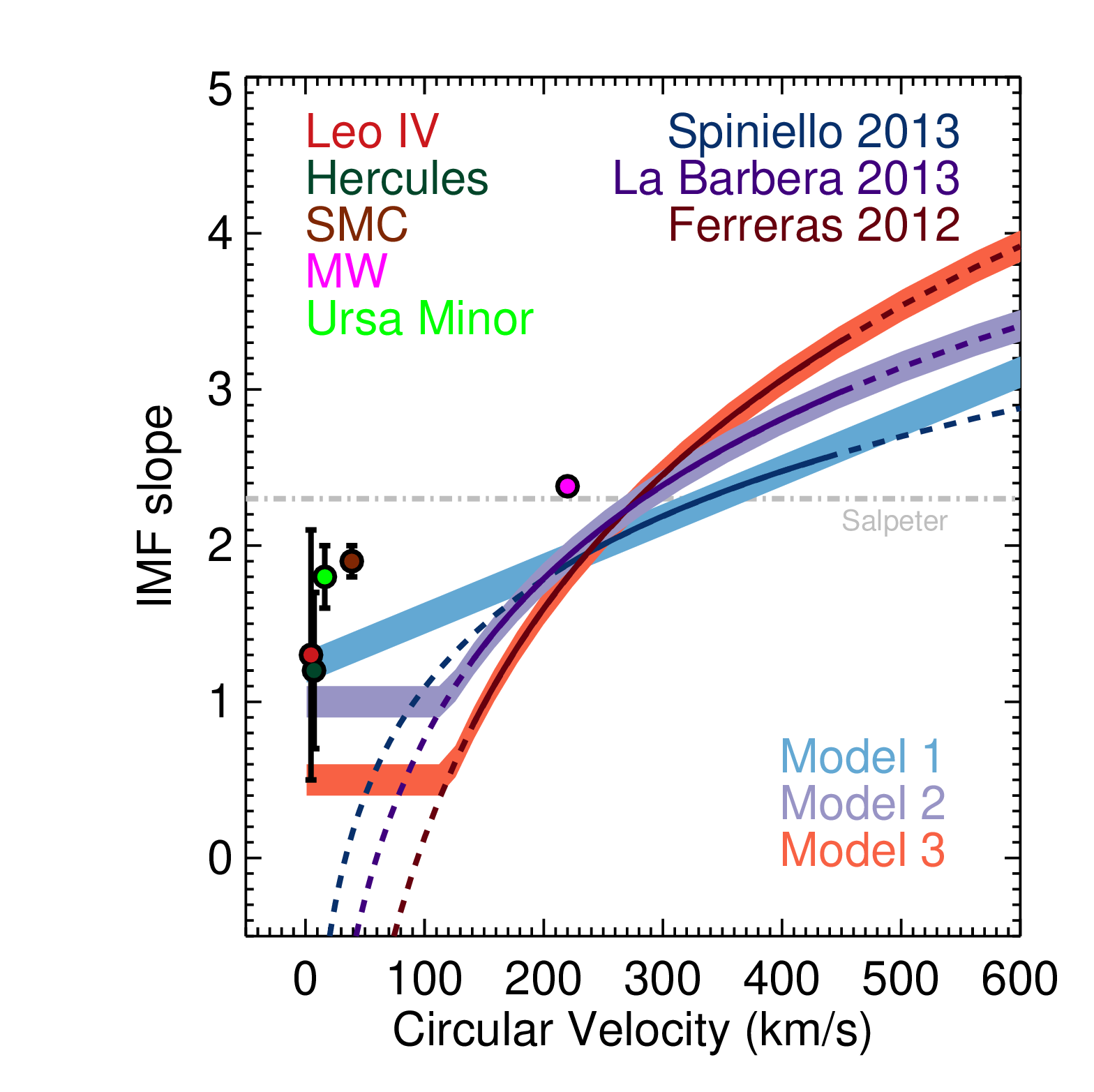}
\caption{The empirical models of circular velocity of the galaxy vs the IMF slope for the unimodal IMF variation case. The blue, purple and orange shaded regions show the models 1, 2, and 3 respectively, which are largely driven by the work of \citet{spiniello2013}, \citet{LaBarbera_2013} and \citet{ferreras_2013_imf}. The lines show the exact relations by these authors, where the solid region is the range of the observational data and the dashed lines show the extrapolation of the relations. The individual points are taken from resolved star counts in nearby galaxies as discussed in \textsection \ref{mod-imf}. A Salpeter IMF has a slope of 2.3 on this scale, and is indicated with the dashed grey line.}
\label{fig_empirical}
\end{figure}
%%%%%%%%%%%%%%%%%%%%%%%%%%%%%%%%%%%%%%%%%%%%%%%%%%%%%%%%%%%%%%%%

When examining IMF variation, most studies assume the IMF is an unbroken power law where the slope is allowed to vary, a so called `unimodal' model. This form is convenient to avoid the complications of arbitrary IMF shapes, however, the exact form of the IMF is poorly constrained. Motivated by the observed turnover in the power law in the local universe, the so-called `bimodal' model, assumes a shallower slope below $<$ 0.6 \Msun. The shape of the IMF below $<$ 0.6 \Msun does not largely affect the broad band stellar light, which is dominated by higher mass stars. For this reason, converting between stellar masses determined by two IMFs which vary only in the low mass slope, as in the case of Kroupa to Salpeter, can largely be done with a single offset for all galaxies regardless of type \citep{pforr_2012}. Nonetheless, we will calculate separate stellar masses for both the unimodal and bimodal IMFs, largely as a consistency check between the proposed forms of IMF variation and the uncertainty associated with SED fitting. 

In addition to the bimodal and unimodal cases, we also calculate stellar masses using a universal \citet{chabrier_imf} IMF, where 
\beq
dN/dm \propto \\
\begin{cases}
  \mathrm{exp}\left[-\frac{(\mathrm{Log}_{10} m - \mathrm{Log}_{10}(0.08))^2}{2(0.69)^2}\right] , & 0.08 > m > 1 M_\odot \\
  m^{-2.3}, & 1 > m > 100 M_\odot 
\end{cases}
\eeq
where $dN/dm$ are the number of stars per interval of mass, and $m$ is the mass. This distribution of masses assumes that there is an upper limit to the mass at which stars form of 100 $M_\odot$ and a lower limit of 0.08 $M_\odot$. This IMF is common in studies of galaxy evolution, and will serve as a reference point against which to measure the effect of IMF variations. In the standard form that we use in this paper, the \citet{chabrier_imf} IMF gives similar stellar distributions to the \citet{kroupa} IMF. However, the piecewise form of the Kroupa IMF does allow for an easier exploration of the possible systematic variation of the low mass slope \citep{Kroupa_review_2013}.

\subsection{Unimodal variation}

The IMF in the case of unimodal variation is similar to a \citet{salpeter_imf}, with a power-law distribution across the allowable range of stars and is written as
\beq
dN/dm \propto m^{\alpha} \hspace{0.5cm} \mathrm{if}\ 0.08 < m < 100 M_\odot 
\label{equ_uni}
\eeq
where $\alpha$ is the IMF slope and is allowed to vary depending on the empirical relations below.

The studies of \citet{ferreras_2013_imf}, \citet{LaBarbera_2013} and \citet{spiniello2013} have each presented empirical relations of unimodal variations of the form: 
\beq
\alpha = a \mathrm{Log}_{10}\left(\frac{\sigma}{200 \mathrm{km/s}}\right) + b
\eeq
where $\alpha$ is the IMF slope and $\sigma$ is the galaxy velocity dispersion. \citet{ferreras_2013_imf} found (a,b) $\Rightarrow$ (4.87, 2.33), while \citet{LaBarbera_2013} found (a,b) $\Rightarrow$ (3.4, 2.30) and \citet{spiniello2013} found (a,b) $\Rightarrow$ (2.3, 2.13). These scalings are shown as the dotted lines in Figure \ref{fig_empirical}, where the circular velocity is assumed to be $\sqrt{2}\sigma$. Unfortunately, the form of this parameterization requires that the IMF slope goes to negative infinity as the velocity dispersion approaches zero. Of course, this parameterization was only used for velocity dispersions of $>$ 100 km/s and are unconstrained by data below this value. 

The principle constraints on the IMF slope in low mass systems comes from a sample of four low mass galaxies where direct star counts were made, as measured and compiled in \citet{Geha2013}. These galaxies, along with the measurement of the Milky Way are shown in Figure \ref{fig_empirical}. The slope and velocity dispersion of Hercules and Leo IV are measured in \citet{Geha2013}, while following those authors we take the SMC IMF results from \citet{Kalirai_mc_imf} and the velocity dispersion from \citet{Harris_smc}, the Ursa Minor IMF slope from \citet{Wyse_ursa_imf} and the velocity dispersion from \citet{wolf2010}, while the Milky Way IMF slope was measured in \citep{Bochanski_mw_imf}. Clearly, these individual galaxies strongly suggest the IMF slope does not asymptote to negative infinity in low mass galaxies.

Given these constraints on the IMF slope, we will use three different models for unimodal IMF variation. Model 1 is assumed to be linear with circular velocity and is designed to include the lowest mass systems as well as the \citep{spiniello2013} results. 
\beq
\alpha = 0.00315 \ v_{\mathrm{circ}} + 1.22 \tag{Model 1}
\eeq
Model 2 and 3 are similar to \citep{LaBarbera_2013} and \citet{ferreras_2013_imf} respectively, but are truncated to have a constant IMF slope in low mass systems. 
\beq
\alpha =
\begin{cases}
3.4 \ \mathrm{Log}_{10}\left(\frac{v_{\mathrm{circ}}}{282 \mathrm{km/s}}\right) + 2.3 & v_{\mathrm{circ}} > 117 \mathrm{km/s} \\
1.0 & v_{\mathrm{circ}} < 117 \mathrm{km/s} \tag{Model 2}
\end{cases}
\eeq

\beq
\alpha = 
\begin{cases}
4.87 \ \mathrm{Log}_{10}\left(\frac{v_{\mathrm{circ}}}{282 \mathrm{km/s}}\right) + 2.33 & v_{\mathrm{circ}} > 119 \mathrm{km/s} \\
0.5 & v_{\mathrm{circ}} < 119 \mathrm{km/s} \tag{Model 3}
\end{cases}
\eeq
All three of these models are shown as shaded regions in Figure \ref{fig_empirical}. We have also assumed that the scatter in the IMF slope at fixed velocity dispersion is $\pm$ 0.1, although the principle effect of increasing the allowed scatter is an increase in the width of the stellar mass probability distribution rather than a change in the median value.

\subsection{Bimodal variation}
The bimodal IMF has a flat distribution at the low mass end and a power law slope at the massive end.
\beq
dN/dm \propto
\begin{cases}
m^{-1.0}, & 0.08 < m < 0.6 M_\odot \\
m^{\alpha}, & 0.6 < m < 100 M_\odot 
\end{cases}
\label{equ_bimo}
\eeq

\citet{LaBarbera_2013} and \citet{ferreras_2013_imf} have presented fits to the IMF variation in the bimodal case, and we use these parameterizations for bimodal analogues to the unimodal models 2 and 3 as such:
\beq
\alpha =
\begin{cases}
5.1 \ \mathrm{Log}_{10}\left(\frac{v_{\mathrm{circ}}}{282 \mathrm{km/s}}\right) + 2.7 & v_{\mathrm{circ}} > 131 \mathrm{km/s} \\
1.0 & v_{\mathrm{circ}} < 131 \mathrm{km/s} \tag{Model 2B}
\end{cases}
\eeq

\beq
\alpha = 
\begin{cases}
7.19 \ \mathrm{Log}_{10}\left(\frac{v_{\mathrm{circ}}}{282 \mathrm{km/s}}\right) + 2.85 & v_{\mathrm{circ}} > 133 \mathrm{km/s} \\
0.5 & v_{\mathrm{circ}} < 133 \mathrm{km/s} \tag{Model 3B}
\end{cases}
\eeq
where we again assume that the scatter in the IMF slope at fixed velocity dispersion is $\pm$ 0.1.

\section{Stellar population synthesis and fitting spectral energy distributions}\label{stelpop}

We will derive galaxy stellar masses by comparing the available $ugriz$ SDSS photometry to large samples of synthesised photometry in which the input parameters sample the full range of galaxy properties. The stellar population synthesis results were calculated with the Flexible Stellar Population Synthesis (FSPS) code \citep{conroy_fspsI} using the BaSeL 3.1 spectral libraries \citep{Lejeune_1997, Lejeune_1998, westera_2002} and isochrones from the Padova group \citep{Marigo_padova_2007, Marigo_padovaII_2007}, which include an updated treatment of thermally pulsating AGB stars. 

We create three sets of synthesized galaxies, each of which has a different treatment of the IMF (Chabrier, bimodal, unimodal), but all the remaining parameters (eg. dust, metallicity, age, star formation histories) are sampled in the same way for each set. The parameter ranges are similar to those of \citet{Salim+07} and \citet{mcgee_11}. In particular, the age of the galaxy is chosen at random such that the distribution is uniform in the logarithm between 0.1 Gyr and the age of the Universe at the observed galaxy redshift. We use the two-component dust model of \citet{Charlot+00} which assumes that young stars have enhanced dust attenuation when compared to older stellar populations. This is thought to arise because the hot, young stars disrupted and evaporate the dust clouds during their lifetime. This model parameterizes the $V$-band optical depth that attenuates young stars ($<$ 10$^7$ years) as $\tau_v$, which we allow to vary between 0 and 6, with a distribution which peaks at $\tau_v$ = 1.2. Old stars ($>$ 10$^7$ years) are attenuated by a modified optical depth, $\mu_v \tau_v$, where $\mu_v$ varies between 0.1 and 1 with a peak near 0.3. The metallicity is allowed to vary between 0.1 and 1.6 of the solar metallicity, which is assumed to be Z = 0.0190.

%%%%%%%%%%%%%%%%%%%%%%%%%%%%%%%%%%%%%%%%%%%%%%%%%%%%%%%%%%%%%%%
\begin{figure}
\leavevmode \epsfysize=9.0cm \epsfbox{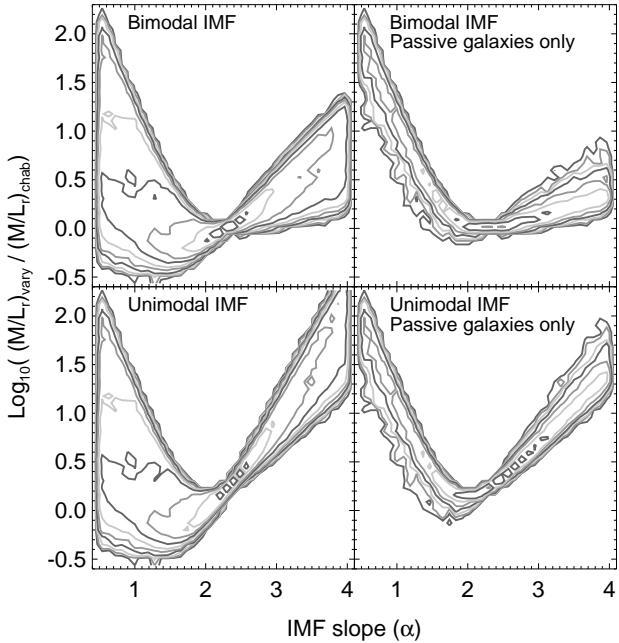}
\caption{The mass to light ratios in the $r$ band for unimodal and bimodal IMF divided by the mass to light ratio of the same synthesised galaxy with a Chabrier IMF. These are shown as a function of $\alpha$ the slope of the massive end of the IMF as defined in equations \ref{equ_uni} and \ref{equ_bimo}. The left plots show the contours for the full range of galaxy parameters while the right panels are restricted to passive galaxies (sSFR $<$ 10$^{-11}$ yr$^{-1}$). }
\label{fig_mlratios}
\end{figure}
%%%%%%%%%%%%%%%%%%%%%%%%%%%%%%%%%%%%%%%%%%%%%%%%%%%%%%%%%%%%%%%%

The star formation history of each galaxy is a combination of a star formation rate which exponentially declines with time and randomly occurring interruptions of bursts of star formation with varying length and strength. The backbone SFR is described by SFR $\propto$ exp($-\gamma t$), where $\gamma$ is uniformly distributed between 0 $\geq$ $\gamma$ $\geq$ 1 Gyr$^{-1}$ and $t$ is the time since the galaxy started forming stars. Bursts of star formation are then interspersed on these star formation histories such that a given galaxy has a 25 per cent change of undergoing a burst within any given Gyr. The bursts last for a randomly distributed time between 30 Myr and 300 Myr and have strengths such that they produce between 0.03 and 4 times the stellar mass the galaxy had formed at the point of burst onset. This burst frequency was advocated by \citet{kauffmann_mass} on the basis of D$_n$(4000) and H$\beta$ strengths. The above distributions of parameters are sampled to create 10$^5$ synthesised galaxy histories. 

For each of these synthesised galaxy spectra, observed magnitudes are made by a convolution with the SDSS filter transmission curves after redshifting to 7 equally spaced redshifts between z=0.02 and z=0.20. In Figure \ref{fig_mlratios}, we show the distribution of M/L depends of the IMF slope ($\alpha$) for the cases of unimodal and bimodal variation. The mass to light ratios are plotted in the $r$ band and are divided by the mass to light ratio of a synthetic model with the same galaxy parameters (eg. age, metallicity, star formation history, etc) but with a Chabrier IMF. In the left panels this is shown for all models, while in the right panels the models are restricted to those currently passive galaxies (with instantaneous specific star formation rates less that 10$^{-11}$ yr$^{-1}$). Notice that in the bimodal model variation, when the IMF slope is 2.3, the ratio of the M/L to that of a Chabrier model is essentially 1, as to be expected. In the unimodal variation model the 2.3 region is greater than one, as this is essentially the ratio of a Salpeter M/L to a Chabrier M/L.  At a given IMF slope, there is up to two orders of magnitude variation in the ratio of the M/L values, which clearly indicates the necessity of a detailed SED fitting of the photometry.

\subsection{SED fitting}

We have now calculated synthesised magnitudes for a wide range of sample galaxies at a range of redshifts and with three different treatments of the IMF. In this section, we will describe how we compare these synthesised fluxes to the observed sample of galaxy fluxes to obtain determinations of the stellar mass. We do this for each observed galaxy by finding the scale factor, $a_i$, which minimizes the $\chi^2_i$ of each model galaxy, $i$, in the following equation.
\begin{equation}
\chi^2_i = \sum_X \left(\frac{F_{{\rm obs},X}-a_i F_{{\rm mod}_i,X}}
{\sigma(F_{{\rm obs},X})}\right)^2 
\label{equ-scaling}
\end{equation}
The sum over $X$ represents each of the 5 bands of SDSS photometry ($u$,$g$,$r$,$i$,$z$). The observed flux in the Xth band is given as $F_{{\rm obs},X}$, while the flux of the model galaxy is written as $F_{{\rm mod}_i,X}$. $\sigma(F_{{\rm obs},X})$ is the error of flux in the each observed band. This fitting is done for each galaxy resulting in $\chi^2$ values for each model. These $\chi^2$ values are used to define a weight w$_i$ = exp($-\chi^2$/2). The probability distribution function of the stellar mass is then created by assigning this weight to the stellar mass of the model galaxy after multiplying by the scale factor $a_i$. These weights are then compounded for all models, and the median value of the PDF is assumed to be the best estimate of the observed galaxy's stellar mass. The 1$\sigma$ error bars were calculated directly from this PDF. 

This procedure is followed for each galaxy in the sample by using the model catalog closest in redshift. The models are directly shifted to the specific redshift of the galaxy. We have added additional uncertainty to the SDSS photometry due to the contribution of zero-point errors of ($u$,$g$,$r$,$i$,$z$) $\Rightarrow$ (0.03, 0.01, 0.01, 0.01, 0.02) magnitudes \citep{uband}

\section{Results}\label{results}

\subsection{Stellar Mass Functions}

\begin{figure}
\leavevmode \epsfysize=9.0cm \epsfbox{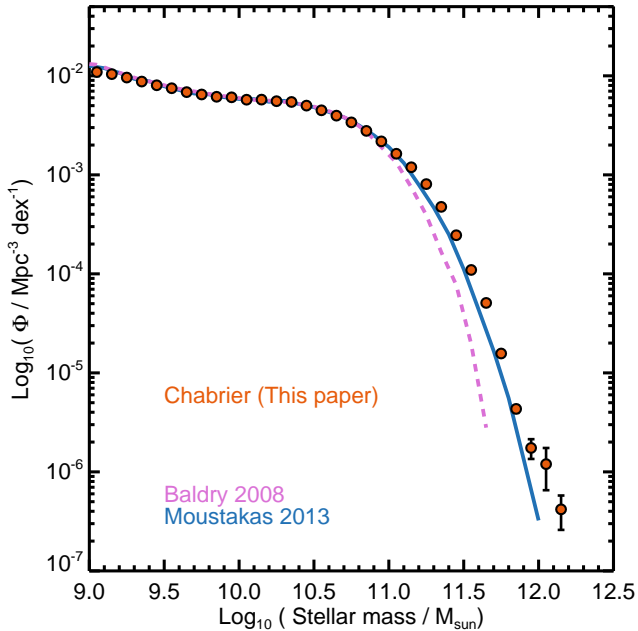}
\caption{The stellar mass function in the local Universe using a Chabrier IMF. The orange points are the results from this paper. The error bars are calculated from Poisson uncertainty only, and except at high masses are smaller than the point size. At the faint end, good agreement is seen with the stellar mass functions of \citet{Baldry_08} and \citet{moustakas_stelmass}. At the massive end, our results agree well with the Moustakas et al. results. The discrepancy with Baldry et al. at the massive end is due to a combination of the smaller volume probed by Baldry et al. and the use of \petromag\ instead of \modelmag.}
\label{fig_chabrier}
\end{figure}

We can now examine how our various models for connecting the IMF to circular velocity has affected the resulting stellar mass function, and thus how we expect the stars to be distributed into galaxies. Given the potential for systematic differences in the creation of stellar mass functions, we first present Figure \ref{fig_chabrier}. This shows the stellar mass function when the SED fitting was done with a stellar population catalog that assumes a universal Chabrier IMF for all galaxies. The stellar mass function is calculated using the V$_{max}$ weighting scheme as discussed in section \ref{sec-vmax}, and results in the number of galaxies per cubic Mpc within each dex of stellar mass as a function of the stellar mass. Our results are shown with the orange points, which include errors calculated assuming only statistical Poisson uncertainty. This uncertainty is generally smaller than the point size.

%%%%%%%%%%%%%%%%%%%%%%%%%%%%%%%%%%%%%%%%%%%%%%%%%%%%%%%%%%%%%%%
%%%%%%%%%%%%%%%%%%%%%%%%%%%%%%%%%%%%%%%%%%%%%%%%%%%%%%%%%%%%%%%%

%%%%%%%%%%%%%%%%%%%%%%%%%%%%%%%%%%%%%%%%%%%%%%%%%%%%%%%%%%%%%%%
\begin{figure*}
\leavevmode \epsfysize=9.0cm \epsfbox{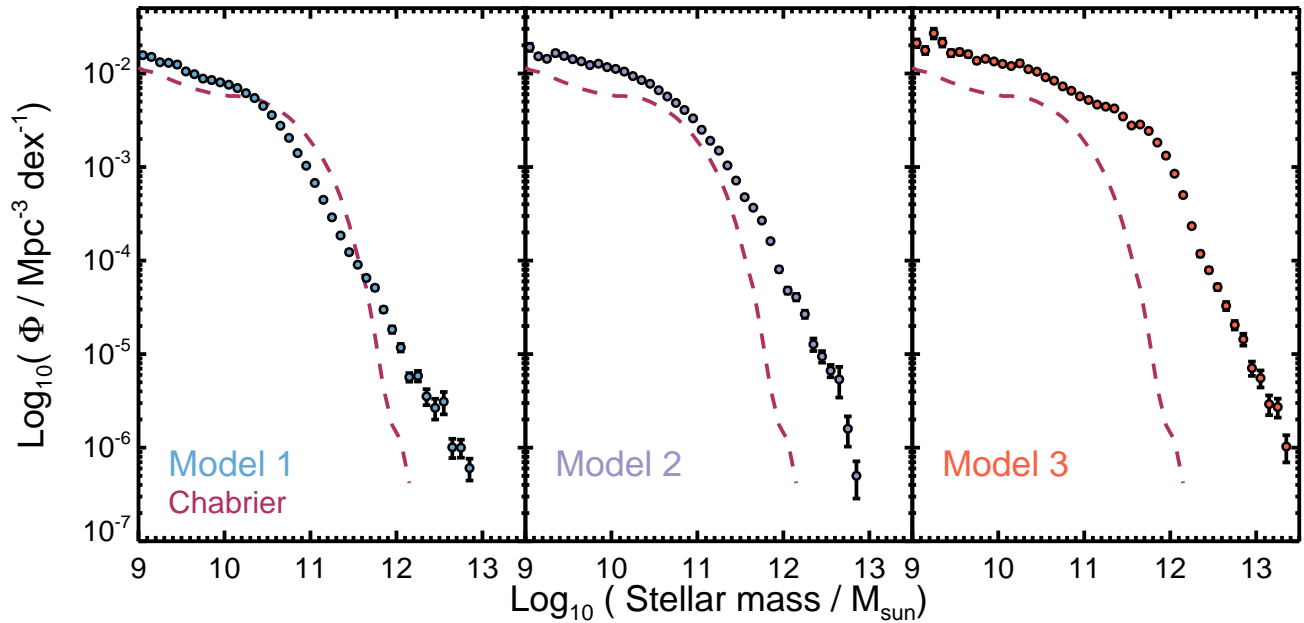}
\caption{The stellar mass functions with three different models for the IMF variation. Each model provides a different scaling between galaxy circular velocity and the IMF slope for the case of unimodal changes in the IMF as discussed in \textsection \ref{mod-imf} and is applied to all galaxies regardless of properties. The dashed magenta line indicates the stellar mass function determined using a Chabrier IMF for the same sample.}
\label{fig_unistelmass}
\end{figure*}
%%%%%%%%%%%%%%%%%%%%%%%%%%%%%%%%%%%%%%%%%%%%%%%%%%%%%%%%%%%%%%%%

%%%%%%%%%%%%%%%%%%%%%%%%%%%%%%%%%%%%%%%%%%%%%%%%%%%%%%%%%%%%%%%%%%%%%%%%%%%%%%%%%%%%%%%
\begin{table*}
\begin{tabular}{lccccc}
\hline 
Model & Log$_{10}$(C) & Log$_{10}$ M$^*$ & $\alpha$ & $\sigma$ & $\beta$  \\ 
\hline
Chabrier &     -2.66 $\pm$      0.01 &     10.40 $\pm$      0.02 &      0.50 $\pm$      0.01 &     -1.22 $\pm$      0.01 &      3.04 $\pm$      0.07 \\ 
Model 1 &     -1.90 $\pm$      0.10 &     10.94 $\pm$      0.02 &     -0.26 $\pm$      0.01 &     -0.86 $\pm$      0.06 &      0.67 $\pm$      0.02 \\ 
Model 2 &     -2.33 $\pm$      0.02 &     10.47 $\pm$      0.02 &     -0.46 $\pm$      0.01 &     -1.13 $\pm$      0.03 &      1.53 $\pm$      0.05 \\ 
Model 3 &     -2.74 $\pm$      0.03 &     11.88 $\pm$      0.02 &      0.30 $\pm$      0.00 &     -1.25 $\pm$      0.02 &      1.30 $\pm$      0.05 \\ 
\hline
\multicolumn{6}{| l |}{Late types: Chabrier}\\
\hline
Model 1 &     -1.69 $\pm$      0.11 &     11.02 $\pm$      0.02 &      0.23 $\pm$      0.00 &     -0.69 $\pm$      0.06 &      0.65 $\pm$      0.02 \\ 
Model 2 &     -1.63 $\pm$      0.11 &     11.02 $\pm$      0.02 &     -0.23 $\pm$      0.00 &     -0.66 $\pm$      0.05 &      0.65 $\pm$      0.02 \\ 
Model 3 &     -2.32 $\pm$      0.05 &     10.83 $\pm$      0.02 &     -0.32 $\pm$      0.01 &     -0.95 $\pm$      0.03 &      0.99 $\pm$      0.03 \\ 
\hline
\end{tabular}
\label{tab_stelfun}

\caption{Table of the parameters describing the best fit stellar mass functions for the generalized Saunders function given in Equation \ref{equ-gen-saunders}. The first four rows give parameters for the stellar mass functions where all galaxies have the same IMF model as shown in Figure \ref{fig_unistelmass}. The final three rows give the parameters for the stellar mass functions when late type galaxies are assumed to have a universal Chabrier IMF, as shown in Figure \ref{fig_unistelmass_early}. }
%\label{tab_stelfun}
\end{table*}
%%%%%%%%%%%%%%%%%%%%%%%%%%%%%%%%%%%%%%%%%%%%%%%%%%%%%%%%%%%%%%%%%%%%%%%%%%%%%%%%%%%%%%%%

The assumption of a universal Chabrier IMF is common in many previous studies and thus we can make a direct comparison with previous determinations of the mass function. The dashed pink line shows the results from \citet{Baldry_08} while the solid blue line shows recent results from \citet{moustakas_stelmass}. At masses below 10$^{11}$ \Msun, our results are essentially indistinguishable from the other studies. However, above this mass there are some discrepancies. The biggest difference is that the Baldry et al. mass function has significantly fewer galaxies at fixed mass at this massive end. This is likely due to a combination of two factors. First, the Baldry et al. results use \petromag\ instead of the \modelmag\ that Moustakas et al and this paper use. It is known that \petromag\ can underestimate the flux in massive galaxies compared to \modelmag. Secondly, Baldry et al. were principally interested in the low mass slope of the mass function, and thus used a catalog which was limited to z $<$ 0.05. This necessarily reduced the volume probed, which predominately affects the rare galaxies at the massive end of the mass function. It is for this reason that the Baldry et al. mass function truncates at $\sim$ 3$\times$10$^{-6}$ Mpc$^{-3}$ dex$^{-1}$, rather than the further order of magnitude that Moustakas et al. and our results probe. Although, this largely explains why the Baldry et al mass function is systematically lower at the massive end, there is still a small discrepancy between our results and Moustakas et al. While the majority of our analysis is similar (eg. redshift range, synthetic galaxy parameters, etc), Moustakas et al. fit 12 bands of photometry from {\sl GALEX} through {\sl WISE} and have a smaller coverage area, largely limited by the need for {\sl GALEX} coverage. Given their increased data, it is not surprising that some differences in the mass determination exist. Indeed, it is perhaps encouraging that our results are so similar despite much smaller wavelength coverage. In the rest of the paper, we will be comparing the stellar mass functions we generate to our mass function with a Chabrier IMF, and thus the differences should be only due to the varying IMF.

\subsubsection{Unimodal IMF variation}

We can now begin to examine the role of the varying IMF. As discussed above, we have determined stellar masses in the unimodal case with three different models, each inspired by a recent determination of the relation between IMF slope and velocity dispersion. We can create stellar mass functions for each of these models using the same V$_{max}$ weightings as in the Chabrier case, because they depend only on the characteristics of the observational selection. However, it is worth remembering that the mass limit depends on the model for IMF variation.

In Figure \ref{fig_unistelmass}, we show the stellar mass function for each of the three models of the IMF in the case of unimodal variation for all galaxies regardless of type. In each panel, the dotted maroon line gives the stellar mass function we found with a universal Chabrier IMF as in Figure \ref{fig_chabrier}. There are several noteworthy features of this figure. First, in a general sense, we see that Model 1 is perhaps the least dramatic, as it crosses the Chabrier MF several times, and for most of the mass range they are within half a dex of each other. In contrast, both Model 2 and Model 3 have more galaxies per volume at every mass compared to the case of a Chabrier IMF. Secondly, we note that the characteristic Schechter shape of the mass function, with an exponential decline at the massive end is largely absent when the IMF varies in this way. Each model still maintains a characteristic break, or `knee', in the mass function but at higher masses is essentially a power law decline. Using the linear regression Bayesian estimator of \citet{kelly_linmixerr} we find that between 10.75 $<$ Log$_{10}$(\Msun) $<$ 12.5, the power law slope is given as $-1.73\pm 0.03$ and $-1.68 \pm 0.06$ for Models 1 and 2.  For Model 3, the slope between 12.0 and 13.5 is steeper, at $-2.17 \pm 0.07$. Thirdly, despite these wholesale changes in the mass function, the low mass slope is very similar in each model, and similar to the slope in the Chabrier MF. 

To better parameterize the stellar mass functions, we have tried several popular forms used in the literature including the Schechter function \citep{Schechter_function}, a modified Schechter function \citep{Sheth_2003} and the Saunders function \citep{Saunders_function}. Unfortunately, none of these parameterizations give reasonable fits to the range of mass functions presented here. As such, we are forced to generalize the Saunders function and fit the following form
\beq\label{equ-gen-saunders}
\phi(M)dM = C \left( \frac{M}{M^*} \right)^\alpha \mathrm{exp}\left[- \frac{1}{2\sigma^2} \mathrm{Log}_{10}^{\beta} \left(1 + \frac{M}{M^*}\right)\right] d\left(\frac{M}{M^*}\right).
\eeq
Notice that when $\beta$ = 2, we recover the Saunders function. The fits were done using \textsc{MPFIT} \citep{mpfit} and are presented in Table \ref{tab_stelfun}. Parameterizations are useful, but with five essentially arbitrary parameters have little physical meaning. Table \ref{tab_stelfun_values_uni_all} gives the individual data points in the stellar mass functions.

%%%%%%%%%%%%%%%%%%%%%%%%%%%%%%%%%%%%%%%%%%%%%%%%%%%%%%%%%%%%%%%%%%%%%%%%%%%%%%%%%%%%%%%
\begin{table*}
\begin{tabular}{lcccc}
\hline 
Stellar Mass & Chabrier & Model 1 & Model 2 & Model 3 \\ 
Log$_{10}$(M/$M_\odot$) & $\phi$/10$^{-5}$ (dex$^{-1}$ Mpc$^{-3}$) & $\phi$/10$^{-5}$ (dex$^{-1}$ Mpc$^{-3}$) &  $\phi$/10$^{-5}$ (dex$^{-1}$ Mpc$^{-3}$) &  $\phi$/10$^{-5}$ (dex$^{-1}$ Mpc$^{-3}$) \\ 
\hline
     9.05 &     1092. $\pm$      71.7 &     1570. $\pm$     131.5 &     1912. $\pm$     255.8 &     2114. $\pm$     293.3 \\ 
     9.15 &     1040. $\pm$      66.2 &     1504. $\pm$     142.9 &     1526. $\pm$     134.0 &     1769. $\pm$     233.5 \\ 
     9.25 &     960.7 $\pm$     62.40 &     1319. $\pm$      91.1 &     1435. $\pm$     115.9 &     2691. $\pm$     444.1 \\ 
     9.35 &     875.4 $\pm$     54.02 &     1300. $\pm$      93.6 &     1652. $\pm$     159.0 &     2147. $\pm$     305.0 \\ 
     9.45 &     802.9 $\pm$     53.83 &     1246. $\pm$     113.3 &     1542. $\pm$     145.4 &     1653. $\pm$     196.4 \\ 
     9.55 &     749.5 $\pm$     48.07 &     1053. $\pm$      68.5 &     1423. $\pm$     102.2 &     1697. $\pm$     180.1 \\ 
     9.65 &     684.7 $\pm$     39.39 &     982.5 $\pm$     62.20 &     1348. $\pm$      94.7 &     1610. $\pm$     169.8 \\ 
     9.75 &     648.1 $\pm$     36.87 &     881.2 $\pm$     50.76 &     1231. $\pm$      80.9 &     1369. $\pm$     121.5 \\ 
     9.85 &     614.3 $\pm$     34.52 &     850.9 $\pm$     52.01 &     1273. $\pm$      79.1 &     1435. $\pm$     126.9 \\ 
     9.95 &     606.7 $\pm$     36.51 &     804.4 $\pm$     48.04 &     1170. $\pm$      70.3 &     1345. $\pm$     110.3 \\ 
    10.05 &     574.0 $\pm$     30.78 &     758.1 $\pm$     41.16 &     1120. $\pm$      65.9 &     1266. $\pm$     104.9 \\ 
    10.15 &     575.2 $\pm$     30.53 &     699.2 $\pm$     39.42 &     1049. $\pm$      61.7 &     1203. $\pm$      84.0 \\ 
    10.25 &     554.5 $\pm$     29.94 &     615.6 $\pm$     32.65 &     938.3 $\pm$     52.86 &     1283. $\pm$     128.9 \\ 
    10.35 &     544.5 $\pm$     31.42 &     546.4 $\pm$     28.53 &     851.1 $\pm$     50.07 &     1111. $\pm$      77.6 \\ 
    10.45 &     500.4 $\pm$     25.57 &     449.5 $\pm$     26.74 &     776.7 $\pm$     46.61 &     1046. $\pm$      71.1 \\ 
    10.55 &     448.5 $\pm$     22.79 &     358.9 $\pm$     18.55 &     662.5 $\pm$     37.04 &     911.5 $\pm$     57.25 \\ 
    10.65 &     396.1 $\pm$     20.39 &     276.5 $\pm$     14.49 &     568.1 $\pm$     32.97 &     838.1 $\pm$     55.66 \\ 
    10.75 &     338.7 $\pm$     16.94 &     204.9 $\pm$     11.33 &     485.0 $\pm$     26.58 &     727.6 $\pm$     55.41 \\ 
    10.85 &     278.0 $\pm$     14.81 &     140.9 $\pm$      7.66 &     407.3 $\pm$     23.12 &     653.5 $\pm$     44.66 \\ 
    10.95 &     217.8 $\pm$     10.89 &     103.3 $\pm$      6.32 &     331.5 $\pm$     19.32 &     570.5 $\pm$     40.46 \\ 
    11.05 &     163.3 $\pm$      8.18 &     67.55 $\pm$      4.02 &     250.3 $\pm$     14.06 &     521.1 $\pm$     34.56 \\ 
    11.15 &     119.3 $\pm$      6.61 &     44.62 $\pm$      2.80 &     191.0 $\pm$     11.05 &     464.9 $\pm$     29.16 \\ 
    11.25 &     80.64 $\pm$      4.15 &     28.99 $\pm$      2.04 &     149.8 $\pm$     12.25 &     442.0 $\pm$     27.06 \\ 
    11.35 &     47.49 $\pm$      2.51 &     18.54 $\pm$      1.30 &     103.9 $\pm$      6.69 &     422.6 $\pm$     33.39 \\ 
    11.45 &     24.55 $\pm$      1.38 &     12.30 $\pm$      1.05 &     71.69 $\pm$      4.64 &     346.9 $\pm$     20.86 \\ 
    11.55 &     10.95 $\pm$      0.70 &     9.037 $\pm$     0.735 &     47.61 $\pm$      3.50 &     277.9 $\pm$     16.42 \\ 
    11.65 &     5.089 $\pm$     0.382 &     6.523 $\pm$     0.532 &     36.78 $\pm$      2.65 &     285.0 $\pm$     17.43 \\ 
    11.75 &     1.569 $\pm$     0.148 &     5.126 $\pm$     0.465 &     26.80 $\pm$      1.96 &     243.1 $\pm$     14.41 \\ 
    11.85 &    0.4327 $\pm$    0.0573 &     2.993 $\pm$     0.299 &     16.13 $\pm$      1.30 &     182.7 $\pm$     10.72 \\ 
    11.95 &    0.1745 $\pm$    0.0473 &     1.833 $\pm$     0.225 &     8.059 $\pm$     0.800 &     132.5 $\pm$      7.84 \\ 
    12.05 &    0.1198 $\pm$    0.0599 &     1.176 $\pm$     0.161 &     4.780 $\pm$     0.561 &     84.89 $\pm$      5.51 \\ 
    12.15 &   0.04185 $\pm$   0.01777 &    0.5674 $\pm$    0.0850 &     4.077 $\pm$     0.509 &     50.22 $\pm$      3.88 \\ 
    12.25 &  --  &    0.5856 $\pm$    0.1028 &     2.675 $\pm$     0.357 &     23.40 $\pm$      1.77 \\ 
    12.35 &  --  &    0.3535 $\pm$    0.0851 &     1.273 $\pm$     0.252 &     11.81 $\pm$      1.23 \\ 
    12.45 &  --  &    0.2666 $\pm$    0.0787 &    0.9439 $\pm$    0.1752 &     7.894 $\pm$     0.785 \\ 
    12.55 &  --  &    0.3105 $\pm$    0.0975 &    0.6659 $\pm$    0.1300 &     5.203 $\pm$     0.590 \\ 
    12.65 &  --  &    0.1010 $\pm$    0.0280 &    0.5375 $\pm$    0.2185 &     3.285 $\pm$     0.474 \\ 
    12.75 &  --  &    0.1000 $\pm$    0.0263 &    0.1594 $\pm$    0.0640 &     2.049 $\pm$     0.306 \\ 
    12.85 &  --  &   0.06047 $\pm$   0.01859 &   0.05004 $\pm$   0.02371 &     1.442 $\pm$     0.274 \\ 
    12.95 &  --  &  --  &  --  &    0.7092 $\pm$    0.1554 \\ 
    13.05 &  --  &  --  &  --  &    0.5544 $\pm$    0.1396 \\ 
    13.15 &  --  &  --  &  --  &    0.2922 $\pm$    0.0826 \\ 
    13.25 &  --  &  --  &  --  &    0.2722 $\pm$    0.0747 \\ 
    13.35 &  --  &  --  &  --  &    0.1028 $\pm$    0.0378 \\ 
\hline
\end{tabular}
\label{tab_stelfun_values_uni_all}

\caption{The tabulated stellar mass functions for each of the three different models of unimodal IMF variation and the universal Chabrier model. The models for IMF variation has been applied to all galaxies regardless of their morphological type. The quoted errors reflect only the Poisson uncertainty.}
%\label{tab_stelfun_values_uni_all}
\end{table*}
%%%%%%%%%%%%%%%%%%%%%%%%%%%%%%%%%%%%%%%%%%%%%%%%%%%%%%%%%%%%%%%%%%%%%%%%%%%%%%%%%%%%%%%%

%%%%%%%%%%%%%%%%%%%%%%%%%%%%%%%%%%%%%%%%%%%%%%%%%%%%%%%%%%%%%%%
\begin{figure*}
\leavevmode \epsfysize=9.0cm \epsfbox{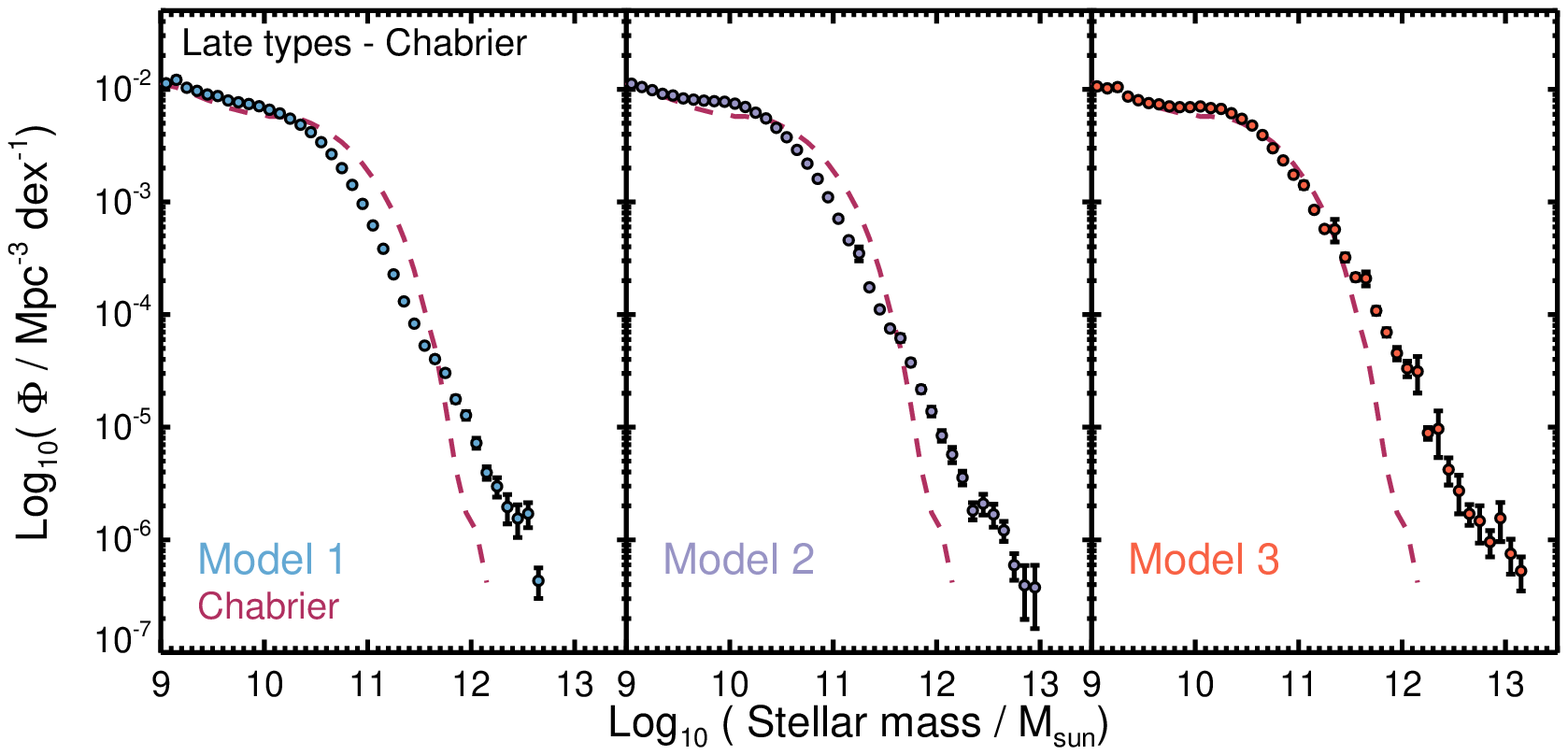}
\caption{The stellar mass functions assuming the IMF varies only for early type galaxies, defined as having a Sersic index greater than 2.5. Late type galaxies are assumed to have a Chabrier IMF. The IMF model for variations are described in \textsection \ref{mod-imf} and the dashed magenta line shows the stellar mass function determined using a Chabrier IMF for all galaxies.}
\label{fig_unistelmass_early}
\end{figure*}
%%%%%%%%%%%%%%%%%%%%%%%%%%%%%%%%%%%%%%%%%%%%%%%%%%%%%%%%%%%%%%%%

As we mentioned in the introduction, the majority of the evidence for IMF variation comes from observations of early type galaxies, so we would like to see the effect of assuming that late type galaxies maintain a universal Chabrier. We define late type galaxies as those with Sersic indicies less than 2.5. Figure \ref{fig_unistelmass_early} shows the results of assuming only the early type galaxies have varying IMFs according to the three models. Notice that the characteristic power law at the massive end of the mass function is largely unchanged in this situation. This is not unexpected, as it is well known that early type galaxies dominate the massive end. This is the most generic feature of each model, regardless of the treatment of late type galaxies. That is, in any reasonable model of IMF variation, the bright end of the mass function becomes a power law, rather than an exponential as in a Schecter function. For each model we find that the slope between 10.75 $<$ Log$_{10}$(\Msun) $<$ 12.5 is $-1.91 \pm 0.04$, $-1.89 \pm 0.03$ and $-1.64 \pm 0.05$  for Models 1, 2, and 3 respectively.  The individual data points for these stellar mass functions are given in Table \ref{tab_stelfun_values_uni_early}.

In this treatment, the low mass end of the mass function is quite similar to that of the universal Chabrier case. Again, this is largely because low mass galaxies are late types, so have the same masses. One additional interesting feature is that the turnover, or the `knee', still occurs earlier in this model than in the Chabrier case. The reason for this is that our models, as shown in Figure \ref{fig_empirical}, have a shallower slope than a Chabrier IMF in systems like our Milky Way, closer to 2.0 than 2.3. Figure \ref{fig_mlratios} shows that IMF slopes of 2.0 naturally lead to lower M/L than those with a slope of 2.3.

%%%%%%%%%%%%%%%%%%%%%%%%%%%%%%%%%%%%%%%%%%%%%%%%%%%%%%%%%%%%%%%%%%%%%%%%%%%%%%%%%%%%%%%
\begin{table*}
\begin{tabular}{lcccc}
\hline 
Stellar Mass & Chabrier & Model 1 & Model 2 & Model 3 \\ 
Log$_{10}$(M/$M_\odot$) & $\phi$/10$^{-5}$ (dex$^{-1}$ Mpc$^{-3}$) & $\phi$/10$^{-5}$ (dex$^{-1}$ Mpc$^{-3}$) &  $\phi$/10$^{-5}$ (dex$^{-1}$ Mpc$^{-3}$) &  $\phi$/10$^{-5}$ (dex$^{-1}$ Mpc$^{-3}$) \\ 
\hline
     9.05 &     1092. $\pm$      71.7 &     1131. $\pm$      73.7 &     1121. $\pm$      77.0 &     1062. $\pm$      76.6 \\ 
     9.15 &     1040. $\pm$      66.2 &     1214. $\pm$     124.4 &     1049. $\pm$      67.3 &     1020. $\pm$      67.2 \\ 
     9.25 &     960.7 $\pm$     62.40 &     1032. $\pm$      66.1 &     983.6 $\pm$     62.78 &     1047. $\pm$     104.4 \\ 
     9.35 &     875.4 $\pm$     54.02 &     968.9 $\pm$     63.57 &     911.5 $\pm$     57.34 &     861.7 $\pm$     54.58 \\ 
     9.45 &     802.9 $\pm$     53.83 &     899.4 $\pm$     58.53 &     882.1 $\pm$     55.64 &     800.3 $\pm$     50.59 \\ 
     9.55 &     749.5 $\pm$     48.07 &     871.8 $\pm$     56.12 &     830.2 $\pm$     53.44 &     754.4 $\pm$     48.79 \\ 
     9.65 &     684.7 $\pm$     39.39 &     796.0 $\pm$     44.41 &     810.6 $\pm$     48.45 &     736.6 $\pm$     50.69 \\ 
     9.75 &     648.1 $\pm$     36.87 &     767.3 $\pm$     42.30 &     794.1 $\pm$     51.54 &     705.8 $\pm$     43.15 \\ 
     9.85 &     614.3 $\pm$     34.52 &     740.7 $\pm$     43.76 &     784.9 $\pm$     43.92 &     694.2 $\pm$     41.08 \\ 
     9.95 &     606.7 $\pm$     36.51 &     707.1 $\pm$     37.74 &     775.9 $\pm$     43.01 &     693.9 $\pm$     39.04 \\ 
    10.05 &     574.0 $\pm$     30.78 &     659.2 $\pm$     34.96 &     746.4 $\pm$     40.92 &     704.7 $\pm$     44.96 \\ 
    10.15 &     575.2 $\pm$     30.53 &     611.2 $\pm$     31.85 &     697.2 $\pm$     37.05 &     681.9 $\pm$     37.09 \\ 
    10.25 &     554.5 $\pm$     29.94 &     548.8 $\pm$     29.02 &     620.3 $\pm$     32.94 &     670.5 $\pm$     38.46 \\ 
    10.35 &     544.5 $\pm$     31.42 &     485.2 $\pm$     24.92 &     551.6 $\pm$     33.32 &     613.0 $\pm$     34.00 \\ 
    10.45 &     500.4 $\pm$     25.57 &     416.4 $\pm$     22.90 &     455.1 $\pm$     24.01 &     546.9 $\pm$     32.31 \\ 
    10.55 &     448.5 $\pm$     22.79 &     339.3 $\pm$     17.43 &     375.3 $\pm$     21.66 &     477.0 $\pm$     26.81 \\ 
    10.65 &     396.1 $\pm$     20.39 &     265.6 $\pm$     14.22 &     289.7 $\pm$     15.96 &     393.0 $\pm$     24.69 \\ 
    10.75 &     338.7 $\pm$     16.94 &     199.5 $\pm$     10.25 &     218.9 $\pm$     12.34 &     299.7 $\pm$     17.58 \\ 
    10.85 &     278.0 $\pm$     14.81 &     141.7 $\pm$      7.34 &     160.2 $\pm$     10.15 &     233.9 $\pm$     15.26 \\ 
    10.95 &     217.8 $\pm$     10.89 &     96.14 $\pm$      5.14 &     110.1 $\pm$      7.22 &     174.5 $\pm$     16.11 \\ 
    11.05 &     163.3 $\pm$      8.18 &     61.90 $\pm$      3.43 &     70.92 $\pm$      4.33 &     140.9 $\pm$     14.61 \\ 
    11.15 &     119.3 $\pm$      6.61 &     38.37 $\pm$      2.21 &     45.74 $\pm$      2.90 &     85.03 $\pm$      6.51 \\ 
    11.25 &     80.64 $\pm$      4.15 &     22.75 $\pm$      1.38 &     34.93 $\pm$      6.52 &     57.55 $\pm$      4.67 \\ 
    11.35 &     47.49 $\pm$      2.51 &     13.09 $\pm$      0.84 &     17.45 $\pm$      1.24 &     57.03 $\pm$     15.53 \\ 
    11.45 &     24.55 $\pm$      1.38 &     8.302 $\pm$     0.607 &     11.11 $\pm$      0.81 &     32.22 $\pm$      3.83 \\ 
    11.55 &     10.95 $\pm$      0.70 &     5.299 $\pm$     0.397 &     7.518 $\pm$     0.645 &     21.47 $\pm$      2.33 \\ 
    11.65 &     5.089 $\pm$     0.382 &     4.026 $\pm$     0.348 &     6.189 $\pm$     0.666 &     20.95 $\pm$      3.91 \\ 
    11.75 &     1.569 $\pm$     0.148 &     3.032 $\pm$     0.268 &     3.751 $\pm$     0.363 &     10.82 $\pm$      1.24 \\ 
    11.85 &    0.4327 $\pm$    0.0573 &     1.771 $\pm$     0.184 &     2.172 $\pm$     0.229 &     6.964 $\pm$     0.846 \\ 
    11.95 &    0.1745 $\pm$    0.0473 &     1.277 $\pm$     0.159 &     1.383 $\pm$     0.188 &     4.538 $\pm$     0.784 \\ 
    12.05 &    0.1198 $\pm$    0.0599 &    0.7238 $\pm$    0.1055 &    0.8433 $\pm$    0.1302 &     3.329 $\pm$     0.679 \\ 
    12.15 &   0.04185 $\pm$   0.01777 &    0.3953 $\pm$    0.0688 &    0.5734 $\pm$    0.1150 &     3.127 $\pm$     1.258 \\ 
    12.25 &  --  &    0.2979 $\pm$    0.0719 &    0.3571 $\pm$    0.0665 &    0.8878 $\pm$    0.1478 \\ 
    12.35 &  --  &    0.1957 $\pm$    0.0658 &    0.1822 $\pm$    0.0399 &    0.9666 $\pm$    0.4724 \\ 
    12.45 &  --  &    0.1548 $\pm$    0.0567 &    0.2099 $\pm$    0.0534 &    0.4202 $\pm$    0.1328 \\ 
    12.55 &  --  &    0.1709 $\pm$    0.0504 &    0.1682 $\pm$    0.0464 &    0.2723 $\pm$    0.1143 \\ 
    12.65 &  --  &   0.04332 $\pm$   0.01505 &    0.1213 $\pm$    0.0299 &    0.1701 $\pm$    0.0434 \\ 
    12.75 &  --  &  --  &   0.05958 $\pm$   0.01841 &    0.1471 $\pm$    0.0603 \\ 
    12.85 &  --  &  --  &   0.03941 $\pm$   0.02151 &   0.09569 $\pm$   0.02930 \\ 
    12.95 &  --  &  --  &   0.03777 $\pm$   0.02320 &    0.1556 $\pm$    0.0661 \\ 
    13.05 &  --  &  --  &  --  &   0.07532 $\pm$   0.02915 \\ 
    13.15 &  --  &  --  &  --  &   0.05292 $\pm$   0.02014 \\ 
\hline
\end{tabular}
\label{tab_stelfun_values_uni_early}

\caption{The tabulated stellar mass functions when IMF variation occurs only in early type galaxies. The values are given for each of the three unimodal IMF varying models as well as the universal Chabrier model. Early type galaxies are those with n$_{\mathrm{Sersic}}$ $>$ 2.5. The quoted erros reflect only the Poisson uncertainty. }
%\label{tab_stelfun_values_uni_early}
\end{table*}
%%%%%%%%%%%%%%%%%%%%%%%%%%%%%%%%%%%%%%%%%%%%%%%%%%%%%%%%%%%%%%%%%%%%%%%%%%%%%%%%%%%%%%%%

\subsubsection{Effect of the shape of the IMF}

\begin{figure}
\leavevmode \epsfysize=6.0cm \epsfbox{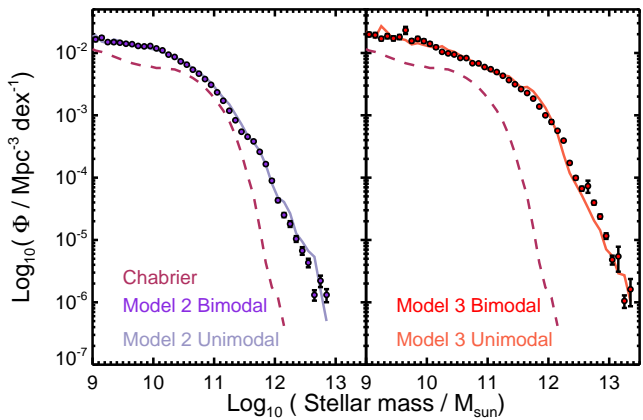}
\caption{The stellar mass functions resulting from a bimodal and unimodal variation of the IMF. Within each parameterization the resulting stellar mass functions are consistent with either form of the IMF. The dashed magenta line shows the stellar mass function determined using a Chabrier IMF for all galaxies.}
\label{fig_bistelmassfun}
\end{figure}

To explore the effect of assuming a different IMF shape we present Figure \ref{fig_bistelmassfun}, which compares the resulting stellar mass functions with a unimodal and bimodal IMF. Both models produce stellar mass functions that are essentially unchanged when compared to those produced with the corresponding parameterization of the unimodal IMF. This was expected, as the break in the IMF slope below 1 \Msun\ does not change the luminosity output significantly. This result shows that the bimodal and unimodal parameterizations of \citet{ferreras_2013_imf} and \citet{LaBarbera_2013} are self consistent. While the shape of the IMF is of significant interest, our results are insensitive to which crude parameterization we use, and thus we use the unimodal variation for the rest of the paper.

%%%%%%%%%%%%%%%%%%%%%%%%%%%%%%%%%%%%%%%%%%%%%%%%%%%%%%%%%%%%%%%
%%%%%%%%%%%%%%%%%%%%%%%%%%%%%%%%%%%%%%%%%%%%%%%%%%%%%%%%%%%%%%%%

\subsection{The correspondence of stellar mass and halos}

\begin{figure*}
\leavevmode \epsfysize=9.0cm \epsfbox{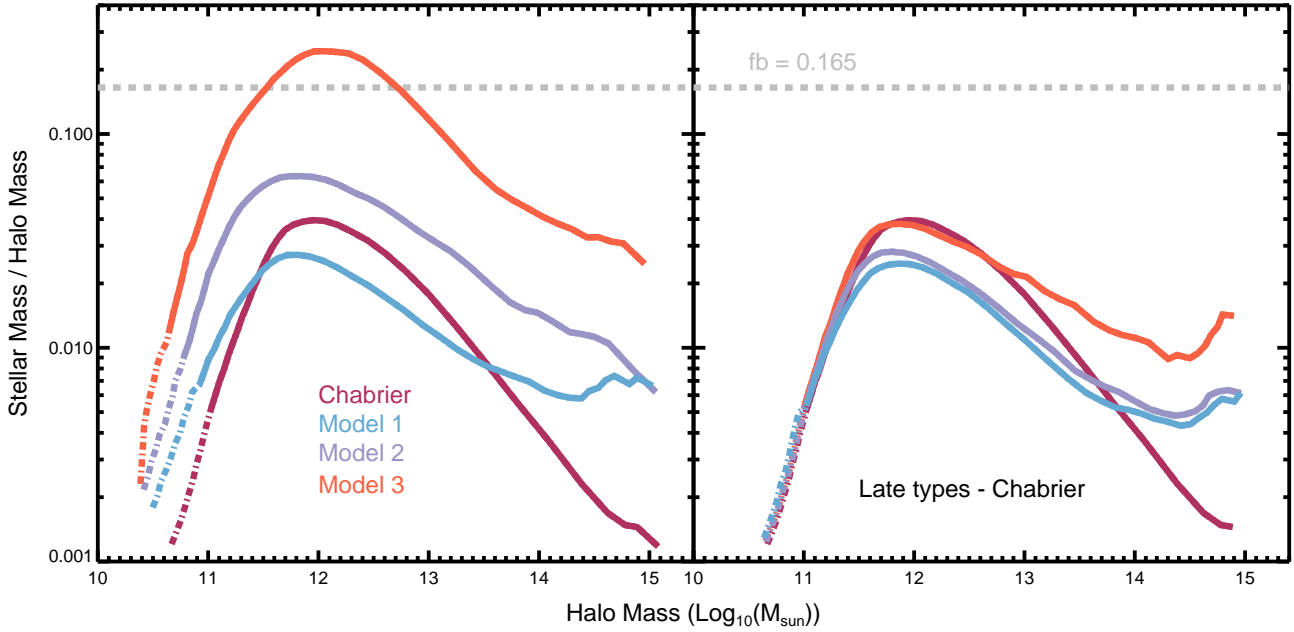}
\caption{The stellar mass to halo mass relation of central galaxies with varying initial mass functions. This is shown as a function of halo masses and is valid for central galaxies. The left panel applies the varying IMF models to all galaxies while the right panel assumes that late type galaxies ($n$ $<$ 2.5) have a Chabrier IMF. These relations are shown for a universal Chabrier IMF as well as the three models described in Section \ref{mod-imf}. The dashed grey line denotes a stellar mass - halo mass ratio of 0.165, equivalent to the universal baryon fraction. }
\label{fig_efficiency}
\end{figure*}

Stellar mass functions alone can be difficult to interpret. That is, given our understanding of the cosmological growth and distribution of dark matter halos, what does a change in the stellar mass function imply for the resultant stellar mass to dark matter ratio in such halos? We make the connection between the halos and the stellar mass by using subhalo abundance matching (SHAM). In this paradigm, it is assumed that each distinct subhalo contains a stellar mass monotonically related to the maximum mass the halo has ever obtained. This is an attempt to correct for satellite galaxies which may have undergone stripping of their dark matter halo and thus seem to have anomalous stellar to dark matter ratios. By making the assignment of galaxies to halos in this manner, we obtain an estimate of how this ratio varies as a function of halo mass for central galaxies. This is presented in Figure \ref{fig_efficiency}. In the left panel, we show the stellar-to-halo mass relation of central galaxies for each of our three IMF variation models applied to all galaxies, while in the right panel we only apply this to the early type galaxies.

%%%%%%%%%%%%%%%%%%%%%%%%%%%%%%%%%%%%%%%%%%%%%%%%%%%%%%%%%%%%%%%
%%%%%%%%%%%%%%%%%%%%%%%%%%%%%%%%%%%%%%%%%%%%%%%%%%%%%%%%%%%%%%%%

%%%%%%%%%%%%%%%%%%%%%%%%%%%%%%%%%%%%%%%%%%%%%%%%%%%%%%%%%%%%%%%%%%%%%%%%%%%%%%%%%%%%%%%
\begin{table*}
\begin{tabular}{lccccc}
\hline 
IMF & \multicolumn{2}{| c |}{Peak Values} & $\alpha$ & $\beta$ & $\rho_*$ \\ 
 & M$_{\mathrm{halo}} (M_\odot)$ & $\frac{M_*}{M_{\mathrm{halo}}}$ & & &Log$_{10}\left(\frac{M_\odot} {\mathrm{Mpc}^{3}}\right)$ \\
\hline
Chabrier &     12.00 &     0.040 &     -0.59 $\pm$      0.02 &      2.04 $\pm$      0.03 &      8.39 $\pm$      0.01 \\ 
Model 1 &     11.80 &     0.027 &     -0.32 $\pm$      0.01 &      1.74 $\pm$      0.08 &      8.29 $\pm$      0.01 \\ 
Model 2 &     11.75 &     0.063 &     -0.37 $\pm$      0.01 &      1.99 $\pm$      0.07 &      8.69 $\pm$      0.01 \\ 
Model 3 &     12.03 &     0.243 &     -0.49 $\pm$      0.01 &      2.30 $\pm$      0.07 &      9.23 $\pm$      0.01 \\ 
\hline
\multicolumn{6}{| l |}{Late types: Chabrier}\\
\hline
Model 1 &     11.89 &     0.025 &     -0.40 $\pm$      0.02 &      1.94 $\pm$      0.05 &      8.23 $\pm$      0.01 \\ 
Model 2 &     11.74 &     0.028 &     -0.38 $\pm$      0.01 &      2.05 $\pm$      0.03 &      8.29 $\pm$      0.02 \\ 
Model 3 &     11.83 &     0.038 &     -0.29 $\pm$      0.01 &      2.09 $\pm$      0.02 &      8.45 $\pm$      0.03 \\ 
\hline
\end{tabular}
\label{tab_main}

\caption{Table of the parameters describing the stellar to halo mass ratio and the integrated stellar density in each model of IMF variation. The columns in the table are (1) the model for IMF variation, (2-3) the halo mass and stellar to halo mass ratio at the peak of the efficiency curve, (4-5) the slope of the high mass end ($\alpha$) and low mass end ($\beta$) of the efficiency curve, (6) the total integrated stellar mass to 10$^{9}$ \Msun. The first four rows have the same IMF model applied to all galaxies, while the last three have a universal Chabrier IMF for all late types and a varying IMF for the early types.}
%\label{tab_main}
\end{table*}
%%%%%%%%%%%%%%%%%%%%%%%%%%%%%%%%%%%%%%%%%%%%%%%%%%%%%%%%%%%%%%%%%%%%%%%%%%%%%%%%%%%%%%%%

First, examining the efficiency curve in the case of a universal Chabrier IMF, we see a characteristic peak in the stellar-to-halo mass ratio near 10$^{12}$ \Msun, with steep declines on either side as seen by many previous studies \citep[eg.,][]{moster_hod, guo_2011_efficiency}. Most generically, it is interesting that a similar shape is seen even with a varying IMF. That is, each model has a distinct peak in the efficiency of galaxy formation in roughly the same region. This did not have to occur, as on either side of this peak we expect the varying IMF to lead to larger mass to light ratios, and potentially a washing out of this peak. 

As a direct result of the similar slope of the massive end power law in the stellar mass function that we previously mentioned, each model with IMF variation results in roughly the same slope of the stellar mass to halo mass relations at high mass. These slopes are tabulated in Table \ref{tab_main}. In contrast, the faint end slope of the relation is only mildly affected.

The primary effect of assuming late type galaxies also have a varying IMF is in the total stellar mass content. Roughly, the different models result in different offsets in the efficiency of galaxy formation but maintain the overall shape of the curve. The gray dashed line in Figure \ref{fig_efficiency} shows the value of the stellar to halo mass ratio if it had the universal baryon fraction (0.165). Our most extreme model, Model 3, exceeds this value at the peak of its efficiency and is clearly disfavoured. 

\subsubsection{The integrated stellar mass in dark matter halos}
%%%%%%%%%%%%%%%%%%%%%%%%%%%%%%%%%%%%%%%%%%%%%%%%%%%%%%%%%%%%%%%
\begin{figure*}
\leavevmode \epsfysize=9.0cm \epsfbox{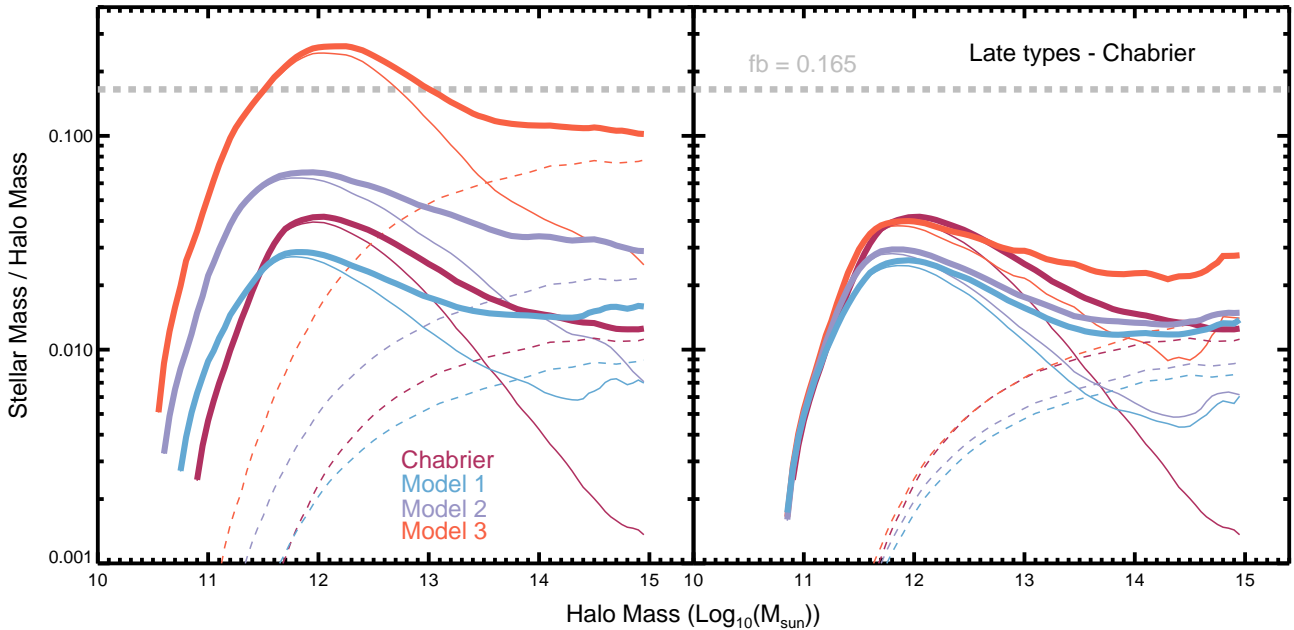}
\caption{The stellar mass to halo mass relation of central (thin solid line), satellite (thin dashed line) and all galaxies (thick solid line) with varying initial mass functions. This is shown as a function of halo mass. The left panel applies the varying IMF models to all galaxies while the right panel assumes late type galaxies have Chabrier IMF. These relations are shown for a universal Chabrier IMF as well as the three models described in Section \ref{mod-imf}. The dashed grey line denotes a stellar mass - halo mass ratio of 0.165, equivalent to the universal baryon fraction. }
\label{fig_efficiency_sats}
\end{figure*}
%%%%%%%%%%%%%%%%%%%%%%%%%%%%%%%%%%%%%%%%%%%%%%%%%%%%%%%%%%%%%%%%

In the SHAM paradigm, when the stellar mass of the galaxy is completely determined by the maximum mass of its dark matter halo, it is straightforward to account for the stellar mass within satellite subhalos. The dark matter simulations follow the trajectory of satellite subhalos and the properties of the large halo they eventually reside in. By summing the stellar mass in these subhalos as a function of host halo mass, we determine the stellar mass to halo mass relation for satellite galaxies. This is presented, along with the central and total stellar to halo mass ratios, in Figure \ref{fig_efficiency_sats}. The left panel of this figure assumes that all galaxies are treated with the same model for IMF variation regardless of their morphological type. We see that the satellite stellar-to-halo mass ratio is steeply rising from $\sim$ 10$^{11}$ \Msun\ to $\sim$ 10$^{12}$ \Msun and then flattens out. This steep rise, which has been seen in many other studies, is a consequence of two factors -- the drop off of the central stellar to halo mass ratio below $\sim$ 10$^{12}$ and the flux limit of the survey. The flux limit of the survey means that the bulk of satellites in $\sim$ 10$^{11}$ \Msun\ halos are not in the sample. 

Interestingly, the point at which the assembled stellar mass in satellite galaxies is greater than that in the central galaxies occurs at 10$^{13.5}$ - 10$^{14}$ \Msun\ regardless of the IMF model and is similar to the value found by \citet{Leauthaud_integrated_2012}. This naturally leads to higher stellar-to-halo mass ratios in groups and clusters, and indeed for Model 1 leads to a nearly flat ratio from 10$^{12}$ \Msun\ and higher. And yet, for groups and clusters, even this value should be treated as a lower limit due to stellar mass that was disrupted and is now in the form of intra-group and cluster stellar mass. This is a result of the implicit assumption in SHAM that when the dark matter halo is disrupted within a more massive halo, the stellar mass within it is also disrupted and distributed throughout the halo. Even without the contribution from ICL, the stellar mass in groups and clusters in Model 3 is near the Universal baryon fraction for that halo. These constraints are even tighter when the contribution to baryons from X-ray gas is added, as we will discus in \textsection \ref{sec-gasmasses}.

In the right panel of Figure \ref{fig_efficiency_sats}, we have assigned subhalo stellar masses assuming the stellar mass function that results when late type galaxies have a universal Chabrier IMF. In effect, this assumes that the morphological fractions of satellites and centrals are the same. Many studies show that this is not true, and thus the satellite stellar mass in this panel should be treated as a lower limit. A higher fraction of early type galaxies would mean more galaxies with bottom heavy IMFs and thus more stellar mass. However, the formation mechanism of satellite early type galaxies is likely different from that of isolated early type galaxies.  In particular, if satellite galaxies have been only recently transformed from late-type to early-type, it may be that more reasonable to assume their IMF is the same as for field late-type galaxies.

\section{Discussion}\label{discuss}

\subsection{Baryon fraction in groups and clusters} \label{sec-gasmasses}

The hot gas in groups and clusters can be detected directly with X-ray telescopes like \Chandra\ and {\sl XMM-Newton}. The total mass and gas mass of these groups and clusters can be determined from the observed surface brightness and temperature profiles \citep{Vikhlinin_2006, sun_large, pratt_rexcess_2009}. However, the X-ray emission depends on the square of the density, and thus falls off rapidly towards the outskirts of clusters. As such the fraction of the mass in X-ray emitting gas is usually measured within R$_{500}$. For our purposes, we would like to extrapolate this result to find the total gas mass with R$_{200}$, which will limit the baryon budget that can be in stellar mass. The ratio M$_{200}$/M$_{500}$ ranges from 1.54 to 1.25 assuming a NFW halo profile with a concentration between 3 and 10. This leads to the strict lower limit on the gas fraction with R$_{200}$, f$_{\mathrm{gas}}$(R$_{200}$), of 0.65 f$_{\mathrm{gas}}$(R$_{500}$). However, the majority of gas profiles show that the gas fraction is rising from the core to R$_{500}$. There are few cases where the X-ray profile is observed to near the virial radius, and these generally show that the gas density rises to the universal density \citep{Bautz_r200_2009, Simionescu_2011, walker_2013}. However, these observations are still at an early stage, and are complicated by the enhanced emission originating from the clumping of gas. Given this, we will assume that the gas fraction within R$_{200}$ is the same as the gas fraction with R$_{500}$, although this is likely an underestimate. 

\begin{figure}
\leavevmode \epsfysize=9.0cm \epsfbox{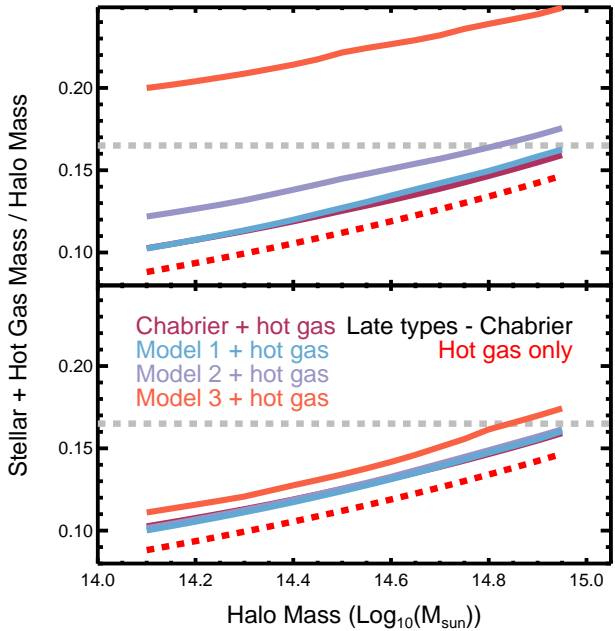}
\caption{The total baryon fraction of groups and clusters with contributions from stellar mass in galaxies and the hot x-ray emitting gas. The top panel shows the results when the stellar content is measured with a varying IMF in all galaxies and the bottom panel has a varying IMF only for early type galaxies. The dashed grey line denotes a stellar mass - halo mass ratio of 0.165, equivalent to the universal baryon fraction. }
\label{fig_baryons}
\end{figure}

The gas fraction is found to be a function of halo mass, such that there are higher gas fractions in more massive systems. \citet{Gonzalez_2013} parameterize this mass dependence as $M_{\mathrm{gas}} = a \left(M_{500}/10^{14} M_\odot \right)^b$ with $a$ = 8.8$\pm$ 0.3 X 10$^{-2}$ and $b$ = 1.26 $\pm$ 0.03. Assuming a M$_{200}$/M$_{500}$ = 1.25, as expected for an extreme c = 10 halo, we can combine this gas mass with the total stellar mass from Figure \ref{fig_efficiency_sats} to find the total baryons in these components. This is presented in Figure \ref{fig_baryons}, where the bottom panel shows the total baryon mass when late type galaxies have a Chabrier IMF, and the top panel has a varying IMF for all galaxies.

As noticed in the central stellar-to-halo mass relation, the most extreme IMF (Model 3 for all galaxies) is ruled out as it requires the halos to have greater than the universal baryon fraction. Interestingly, even the less extreme Model 2 appears to rise above the universal baryon fraction at high masses. The remaining models all tend towards having the universal baryon fraction in the most massive clusters. This is not unexpected as these systems are likely `closed'. Notably, because of the relatively flat stellar mass fractions and the steeply rising x-ray mass fractions, all models have a steep trend in the baryon content. Thus, unless galaxy groups have a significant baryon fraction outside of galaxies and the hot intracluster medium, then some process likely removed or expelled baryons from them, perhaps through AGN feedback \citep{MccarthyAGN, Mccarthy_explusion_2011}. One possibility is that a significant fraction of stars occur in the intragroup medium, but not in the intracluster medium. Although there is a wide range in ICL/IGL measurements, there are none which are several times the mass within galaxies, as would be required to make 10$^{14.1}$ \Msun halos contain the universal baryon fraction \citep{Gonzalez_ICL_2005, zibetti, mcgee_supernova, Budzynski_similar_2013}.

%%%%%%%%%%%%%%%%%%%%%%%%%%%%%%%%%%%%%%%%%%%%%%%%%%%%%%%%%%%%%%%
%%%%%%%%%%%%%%%%%%%%%%%%%%%%%%%%%%%%%%%%%%%%%%%%%%%%%%%%%%%%%%%%

\subsection{Implications for feedback models}

The shape of the efficiency curve has often been taken as evidence that two independent mechanisms inhibit the rapid formation of stars. At the low mass end, the steep increase with halo mass is thought to be related to a star formation specific mechanism, like supernova feedback. In contrast, the rapid decline at high masses is taken asindirect evidence for feedback from supermassive black holes. In the context of our results, a varying initial mass function suggests that AGN feedback has to remove less gas, and not induce such an extreme exponential decline in the number density of high mass galaxies. Ultimately, as the couplings of the AGN to the surrounding media are significantly uncertain, it is unlikely that a major adjustment of models would be required to accommodate IMF variations of the type explored here. For instance, many `sub-grid' models of AGN feedback are targeted to match black hole scaling relations, such as the M$_{\bullet}$ - $\sigma$ relation. As the stellar mass does not enter this particular relation, the accretion of gas onto the black hole would be unaffected. With a varying IMF, there would be more stellar mass formed for a given black hole mass, which would indicate that either the radiative efficiency of the BH accretion disks would be lower than the normally assumed 0.1, or the fraction of that radiated energy that couples to the gas is lower. Both of these model parameters are sufficiently uncertain that such adjustments are expected to be within the bounds of possible models acceptable with the current parameterizations. Clearly, however, as the fraction of radiated energy which couples to the gas decreases the relative importance of AGN feedback decreases as well. In the extreme Model 3, the brightest cluster galaxies have nearly half the baryons locked up in stars, which would require essentially no AGN feedback. However, at the massive end, the observed old populations still require there to be some form of feedback, likely in the form of `radio-mode' feedback which operates in low accretion systems \citep{croton, Bowermodel}. In detail, the effect these results have on the required AGN feedback will depend on the physical mechanism that causes the IMF variation. 

Interestingly, we found that extending the varying IMF to all galaxies could systematically increase the stellar mass density in the Universe. However, the shape of the efficiency curve does not significantly change. This suggests that while feedback continues to be effective in the same mass ranges, so for instance, the scaling of wind speed with halo mass in supernova feedback models need not be altered drastically. Again, an adjustment in the coupling of radiation to gas could suffice as this leads to an increase in the overall star formation efficiency in galaxies.

\subsection{The consistency of IMF variation}

Without a theory of the cause of IMF variation there are still many questions which remain to be explored. For instance, a significant channel of massive galaxy growth occurs through accreting less massive galaxies \citep{vanDokkum_massive_growth_2010}. However, if the relation between IMF slope and velocity dispersion is fundamental, the accreted less massive galaxies must have formed with a shallower IMF slope. This implies that the IMF for {\it in situ} stars in massive galaxies must have been even more extreme to result in IMF-$\sigma$ relation we observe today. In contrast, if the IMF variation is strictly related to metallicity, then the accreted lower mass galaxies may have already been polluted by outflows of the massive galaxy and also formed with a variable IMF. Probing the radial gradients of IMF sensitive features will be a strong constraint on the physical mechanism of this variation. 

As discussed in the introduction, some analysis now finds that the integrated star formation history is equal to the measured stellar density of the universe \citep{Sobral_2013, Behroozi_2013}, in contrast to previous results \citep{Wilkins_2008_imf, Wilkins_2008}. Nonetheless, the corrections for dust extinction and IMF variations in the star formation calibration could still accomadate IMF variations. Indeed, a model where the IMF varies as the Jeans mass goes some way to explain the previously discrepant results \citep{Narayanan_jeans_imf_2012}. However, these integral constraints may indicate that there is variation in the IMF even within a galaxy. 

Similar, integrated constraints, exist from the production of metals. As the majority of metals produced in the universe occur in massive stars, a strongly bottom-heavy IMF has difficulty in producing the metal enrichment observed \citep{weidner_2013}.  However, since the metallicity constraints depend on the enrichment and mass of outflowing gas, this may suggest that massive galaxies had an early `top-heavy' formation event and a longer history of bottom-heavy star formation \citep{weidner_2013}. This also is consistent with an IMF related to the Jeans mass \citep{Narayanan_jeans_imf_2013}.

\section{Conclusions}\label{conclusions}
We have presented an exploration of the effect of IMF variations on the stellar mass function and the distribution of stellar mass in dark matter halos. Guided by recent observations of a systematic variation of the IMF slope in early type galaxies, we have explicitly linked the IMF slope to the circular velocity of the system as calculated from the Fundamental Plane and Tully Fisher relations. We then generated stellar masses for each parameterization by comparing the SDSS photometry to large grids of synthetic stellar populations models which sample age, metallicity, dust and star formation histories. The resulting stellar masses were used to generate stellar mass functions and to examine the stellar to halo mass ratio as a function of halo mass through sub-halo abundance matching. Our conclusions are as follows:

\begin{itemize}
\item The massive end of the stellar mass function resulting from a varying IMF is a power law, instead of the familiar Schechter-like exponential decline. These massive galaxies are dominated by early types and thus this result is insensitive to whether the IMF is allowed to vary in all galaxies or only in early types.
\item The general characteristic shape of the central stellar to halo mass ratio as function of halo mass is robust regardless of IMF model. There is a peak in the ratio at $\sim$ 10$^{12}$ \Msun\ regardless of the exact parameterization or if late types are also allowed to have varying IMFs.
\item The inclusion of IMF variation in late type galaxies largely results in the variation of the total integrated stellar density, rather than a systematic change in the shape of the efficiency curve. Indeed, our most extreme IMF model can not apply in both early and late type galaxies, as it requires some galaxies to have more than their halo's baryon fraction worth of stars. 
\item The power law at the massive end of the stellar mass function with a varying IMF results in a shallower drop off in the stellar-to-halo mass ratio than occurs with a universal IMF. It appears that feedback at this end does not need to be as dramatic as previously thought.
\item The contribution of satellites to the stellar mass of dark matter halos rapidly increases with system mass, such that the total stellar mass to halo mass ratio in groups and clusters is nearly constant. This result occurs for all models of IMF variation. Including the contribution to the baryon budget of hot X-ray gas means the most massive clusters have nearly the universal baryon fraction. Regardless of the IMF model, galaxy groups still have less than the universal baryon fraction in X-ray gas and stellar mass within galaxies.

\end{itemize}

The effect of IMF variations on the cosmic evolution of galaxy properties is of considerable importance for our understanding of galaxy evolution. However, without a good model for the physical process which drives the IMF variation we are limited to imposing empirical relations derived from subsets of the full galaxy population. The advent of large, homogeneous surveys has often led to observational results being presented in terms of their stellar mass, whether it is stellar mass functions or the specific star formation rates of galaxies. In principle, this allows a more physical interpretation of the results, however, it also leads to the underestimate of the systematic uncertainties of such results from sources such as a varying IMF.

\section*{Acknowledgements}

SLM acknowledges support from an NWO grant to George Miley. SLM and RG thank the Leiden/ESA astrophysics program for summer students (LEAPS) which supported RG in Leiden. MLB acknowledges support from an NWO visitor grant and an NSERC discovery grant.

\bibliography{ms}

\end{document}